\documentclass{aa}
\usepackage[varg]{txfonts}

\usepackage{graphicx,array}   
\usepackage{amsmath}
\usepackage{amssymb}    
\usepackage{xspace}    
\usepackage{natbib}
\usepackage{hyperref} 
\hypersetup{
    colorlinks=true,
    allcolors=blue,
    }
\bibpunct{(}{)}{;}{a}{}{,}

\newcommand{\cii}{[C\,{\sc ii}]\xspace}
\newcommand{\oiii}{[O\,{\sc iii}]\xspace}
\newcommand{\cofe}{[C\,{\sc ii}]\(\,158\mu\)m\xspace}
\newcommand{\oee}{[O\,{\sc iii}]\(\,88\mu\)m\xspace}

\newcommand{\ci}{[C\,{\sc i}]\xspace}
\newcommand{\nodata}{\centering\arraybackslash --}
\def\Vhrulefill{\leavevmode\leaders\hrule height 0.7ex depth \dimexpr0.4pt-0.7ex\hfill\kern0pt}

\begin{document}

\title{Early galaxy evolution: The complex interstellar medium distribution
of the $z\sim7$ galaxy A1689-zD1} 
\titlerunning{Early galaxy evolution: The complex ISM distribution
of the $z\sim7$ galaxy A1689-zD1} 

\author{Kirsten K. Knudsen\inst{\ref{inst1}}\thanks{E-mail: kirsten.knudsen@chalmers.se}
\and Darach Watson\inst{\ref{inst2},\ref{inst3}}
\and Johan Richard\inst{\ref{inst4}}
\and David T. Frayer\inst{\ref{inst5}}
\and Seiji Fujimoto\inst{\ref{inst2},\ref{inst3},\ref{inst6}}
\and Hollis Akins\inst{\ref{inst2},\ref{inst3},\ref{inst6}}
\and \\Tom Bakx\inst{\ref{inst1}}
\and Nina Bonaventura\inst{\ref{inst2},\ref{inst3}}
\and Gabriel Brammer\inst{\ref{inst2},\ref{inst3}}
\and Lise Christensen\inst{\ref{inst2},\ref{inst3}}
\and Takuya Hashimoto\inst{\ref{inst7}}
\and Akio K. Inoue\inst{\ref{inst8},\ref{inst9}}
\and Hiroshi Matsuo\inst{\ref{inst10},\ref{inst11}} 
\and Micha\l\  J.\ Micha\l owski\inst{\ref{inst12}}  
\and Jorge A.\ Zavala\inst{\ref{inst10}}
}

\institute{
Department of Space, Earth and Environment, Chalmers University of Technology, SE-412 96 Gothenburg, Sweden\label{inst1} \and 
Cosmic Dawn Center (DAWN), University of Copenhagen, 2200 Copenhagen N, Denmark\label{inst2} 
\and
Niels Bohr Institute, University of Copenhagen,
Jagtvej 128, DK-2200 Copenhagen N, Denmark\label{inst3} 
\and
Univ Lyon, Univ Lyon 1, Ens de Lyon, CNRS, Centre de Recherche
Astrophysique de Lyon UMR5574, F-69230, Saint-Genis-Laval, France \label{inst4}
\and
Green Bank Observatory, P.O. Box 2, Green Bank, WV 24944,
USA\label{inst5}\and
Department of Astronomy, The University of Texas at Austin, Austin, TX 78712\label{inst6}\and
Tomonaga Center for the History of the Universe (TCHoU), Faculty of Pure and Applied Sciences, University of Tsukuba, Tsukuba, \\ $^{  }$ Ibaraki 305-8571, Japan\label{inst7}\and
Department of Physics, School of Advanced Science and Engineering, Faculty of Science and Engineering, Waseda University, 3-4-1 Okubo, Shinjuku, Tokyo 169-8555, Japan\label{inst8}\and
Waseda Research Institute for Science and Engineering, Faculty of Science and Engineering, Waseda University, 3-4-1 Okubo,
Shinjuku, Tokyo 169-8555, Japan\label{inst9}\and
National Astronomical Observatory of Japan,
2-21-1 Osawa, Mitaka, Tokyo 181-8588, Japan\label{inst10}\and
Graduate University for Advanced Studies (SOKENDAI), 2-21-1 Osawa,
Mitaka, Tokyo 181-8588, Japan\label{inst11}\and 
Astronomical Observatory Institute, Faculty of Physics and Astronomy, Adam Mickiewicz University, Poznań, Poland \label{inst12}
}

\authorrunning{Knudsen et al.} 

\date{Received 2 December 2024 / Accepted 30 June 2025} 

\abstract
{We observed the gravitationally lensed ($\mu = 9.6\pm0.19$) galaxy A1689-zD1
at $z = 7.1$ in bands 3, 6, and 8 of the Atacama Large
Millimeter/submillimeter Array. These high-resolution observations ($\approx
200$\,pc) enabled us to separate the source into five components in the \cofe
and \oee emission lines within a projected distance of 2\,kpc. Even though
these components appear to vary strongly from one another in both their line,
continuum, and optical characteristics, the assembly of components do not
show ordered rotation and appear consistent with simulations of a galaxy
system undergoing the process of assembly. The total dynamical mass of the
galaxy ($2\times10^{10}$\,M$_\odot$) is an order of magnitude larger than the
spectrally estimated stellar mass, suggesting a near-complete optical
obscuration of the bulk of the stellar component. Comparing the line ratios
as well as the line properties to other properties such as the star formation
rate, we find that A1689-zD1 is consistent with the relations derived from
local star-forming galaxies. Even though A1689-zD1 lies on local star
formation scaling relations and has a high dust and stellar mass estimate,
the kinematics suggest it is in an early assembly stage, which could lead to
it becoming a disk galaxy at a later stage.}

\keywords{Galaxies: high-redshift -- Galaxies: evolution -- Galaxies: ISM --
Galaxies: individual: A1689-zD1 -- Submillimeter: galaxies}
\maketitle

\section{Introduction}

Galaxies at redshifts of $z>6$ represent the early stages of galaxy formation and evolution, often referred to as cosmic dawn. Discovering these galaxies and determining their physical properties is key to understanding the further phases of galaxy evolution \citep[e.g.,][]{bouwens22b,finkelstein23,rieke23}. Furthermore, these galaxies are essential for understanding the galaxy-scale processes that gave rise to the epoch of reionization \citep[e.g.,][]{hutter21}.   
The gradual change of a steeper slope of the faint end of the ultraviolet (UV) luminosity function with increasing redshift \citep[e.g.,][]{mclure13,finkelstein15,oesch18,bouwens22a} implies an increased relative importance of the sub-$L*$ population's contribution to the overall production of the UV light responsible for  reionization.  
However, mapping galaxies at redshifts $z>6$ is challenged by the enormous
distances resulting in faint appearance,
the redshifting of the
rest-frame UV/optical/near-infrared (NIR) light into wavelengths with low atmospheric transmission, and the varying morphologies (i.e., clumpiness; e.g., \citealt{gullberg18}) of galaxies in the early Universe might be more difficult to resolve with sensitive interferometers.

An efficient way to study the sub-$L*$ $z>6$ galaxy population is through
gravitational lens studies.  The gravitational lensing magnification aids the
time expensive observations of the low-luminosity galaxy population enabling
detailed studies of individual sources.  Using submillimeter wavelength
observations to resolve emission lines is a necessary means for determining
physical properties and characterizing the dynamical state of such galaxies.
The far-infrared (FIR) fine-structure lines of ionized carbon and oxygen has
proven essential for this \citep[e.g.,][]{hashimoto19,harikane20}.  For local
galaxies, even dwarf galaxies, relations between the star formation rate
(SFR) and the line luminosity of the FIR fine-structures lines have been
established \citep[e.g.,][]{delooze14,cormier15}. However, the growing number
of \cii observations of $z>6$ sub-$L*$ galaxies have yielded a more
complicated result for the early epoch
\citep[e.g.,][]{capak15,knudsen16,bradac17,carniani17,hashimoto18,tamura19,laporte19,bakx20,jolly21,fujimoto21,molyneux22};
 for example, some of those early \cii observations yielded several non-detections or line luminosities lower than initially predicted.
With the combination of high sensitivity and sub-arcsecond resolution of modern submillimeter interferometric observatories, it has become possible to resolve line emission even for $z>6$ galaxies.  This makes it possible to conduct dynamical studies \citep[e.g.,][]{carniani13,litke19,spilker22,rowland24}; however, without gravitational lensing, reaching sub-kpc scales remains challenging.
Numerical zoom-in simulations of $z>6$ galaxies show that the kinematics can be very complex. For example, \citet{kohandel19} demonstrate that the same galaxy over the course of hundred million years can change appearance from a merging system to a rotating disk. 
The same simulations, as well as those of \citet{katz22}, also reveal how the predicted observations are dependent on the orientation, showing the difficulty of distinguishing different dynamical states with only low angular resolution. 

The galaxy A1689-zD1 was reported as a $z>7$ candidate with a photometric
redshift of $z = 7.6$ \citep{bradley08} and this was confirmed with X-shooter
spectroscopy that showed a break in the continuum emission at a redshift of $z=7.5\pm0.2$ \citep{watson15}.   
A1689-zD1 was also the first sub-$L*$
with a dust-detection with the Atacama Large Millimeter/submillimeter Array
(ALMA), revealing not only a dust mass similar to that of the Milky Way
interstellar medium (ISM), but also a complex distribution that could be
interpreted as a potentially interacting system \citep{watson15,knudsen17}.  

Some of the brightest FIR fine-structure lines are \cii\,158$\mu$m and \oiii\,88$\mu$m in star-forming galaxies. As carbon has an ionization potential of 11.2\,eV, which is below that of hydrogen, the \cii line is used for both tracing neutral and ionized gas. Ionized oxygen has an ionization potential of 35.1\,eV and, thus, it is only observed in the ionized medium (e.g., H\,{\sc ii} regions).  Thus, while each of these two bright emission lines can be used to resolve the properties of the ISM, when used in combination, they also provide insights to the ionization state of the ISM in early galaxies \citep[e.g.,][]{harikane20,killi23}. 

Here, we present deep imaging of the two FIR fine-structure lines
\cii\,158$\mu$m and \oiii\,88$\mu$m for A1689-zD1. Some of these data were recently  analyzed and the results have begun shedding light on different aspects about FIR properties \citep{bakx21,akins22,Wong2022,killi23}. 
While the original goal of
the \cii observations was to determine the redshift, the \cii and \oiii data enable 
imaging of sub-kpc scales thanks to the gravitational lensing amplification (thereby enabling multi-component decomposition).  
Furthermore, we present detection limits from molecular gas observations. 
The paper is structured as follows. Section~\ref{sect:obs} presents the ALMA
observations and data, along with Green Bank Telescope (GBT) observations of CO(3--2).  Section~\ref{sect:res} gives the results including an analysis of the complex gas kinematics. 
A discussion is given in Sect.~\ref{sect:disc} and our conclusions are summarized in
Sect.~\ref{sect:concl}. We assume a $\Lambda$ cold dark matter cosmology, with $H_0~=~67.7$\,km\,s$^{-1}$\,Mpc$^{-1}$, $\Omega_M = 0.307$, and $\Omega_\Lambda =
0.693$ \citep{planck15}.

\section{Observations}
\label{sect:obs}

\subsection{ALMA data}
Observations of A1689-zD1 were obtained with ALMA to search for the \cofe and subsequently to follow-up the
\oee line (ALMA projects 2015.1.01406.S and 2017.1.00775.S). The \cii observations were tuned to match frequencies
of candidate emission lines found in previous data \citep{knudsen17}. 
The receivers were tuned to 217.19--220.94 and 231.98--235.74\,GHz, where the
line was predicted to be in the former frequency range. 
The calibrators used were J1337$-$1257 for flux calibration, J1256$-$0547 for bandpass, and J1319$-$0049 for gain calibration.   
Data were calibrated
using the ALMA pipeline; we verified the observatory delivered
calibration with only the need for limited additional flagging, however,
three out of ten executions needed the absolute flux calibration redone as
the pipeline had not included the correct values for the flux calibrator. 
After this correction, the uncertainty of the absolute
flux calibration is estimated to be about 5\% \footnote{In line with the technical handbook: \url{https://almascience.eso.org/
documents-and-tools/cycle10/alma-technical-handbook}}.  
Combining the data from all ten executions, we reached an overall depth of
0.18--0.2\,mJy/beam in a 5\,km\,s$^{-1}$ channel\footnote{
The range in rms reflects the variation between different channels of the data cube.}
when imaging using a Briggs weighting 
with a robust parameter of 0.5; the angular resolution is
$0.29''\times0.27''$ and the position angle (PA) is $-80.8^\circ$.  
When combining the lower frequency band with the
continuum data from \citet{knudsen17}, we obtained a root mean square (rms) flux uncertainty of about
6.3\,$\mu$Jy/beam.  

The emission line in the upper frequency band was detected at $\nu = 233.3$\,GHz
and identified as the \cofe emission line corresponding to a redshift of
$z_{\rm [CII]} = 7.133$. Using this redshift ALMA band 8 observations
were obtained for the \oee line. 
The receivers were tuned to 404.40--408.21 and 416.33--420.09\,GHz. 
Calibration was performed using J1229+0203 for flux and bandpass calibration, and J1256$-$0547 for gain calibration.   
The observatory-delivered calibrated data were checked and considered
sufficient, and no further treatment was done.  The data were imaged, 
yielding a continuum map with rms of 29\,$\mu$Jy/beam and an line cube with
an rms of 0.39--0.44\,mJy/beam in $\sim$11\,km\,s$^{-1}$ (15.625\,MHz) channels.  
The resulting images have an angular resolution 
of $0.33''\times0.28''$, PA=\(-87.3^\circ\), which is similar to the
resolution of \cii and band 6 maps. A conservative estimate on the absolute flux calibration is 10\%. A continuum subtraction was performed for both the \cii\ and \oiii\ data cubes using the CASA task {\sc uvcontsub} where the continuum was estimated using the line-free channels.

ALMA band 3 observations of the CO(6--5) and CO(7--6) lines were obtained using a receiver setup of 84.07--87.83\,GHz and 96.16--99.94\,GHz 
(ALMA project 2017.1.00775.S).
The calibrator J1256$-$0547 was used for flux and bandpass calibration, while J1319$-$0049 was used for gain calibration. The observatory-delivered, calibrated data were of sufficient quality. The data were imaged using natural weighting and also using {\it uv}-tapering with a circular Gaussian profile of 0.5\,arcsec resulting in an angular resolution of $0.92''\times0.80''$ (PA=64$^\circ$). The rms in the continuum map is 11\,$\mu$Jy/beam, and in the spectral cube near the frequency of redshifted CO(6--5) [CO(7--6)] the rms is 0.23 [0.20]\,mJy/beam in a 15.625\,MHz channel (corresponding to a 55 [47]\,km\,s\(^{-1}\) channel width). 

\subsection{GBT data}

Using the GBT\footnote{The Green Bank Observatory is a facility of the National Science Foundation operated under a cooperative agreement by Associated Universities, Inc.} observations, we carried out a search for the redshifted CO(3--2) emission with the Q-band (40--50\,GHz) receiver.  Six observing sessions (AGBT14A\_565) were undertaken in September-October 2014 and an additional eight sessions (AGBT18A\_230) were carried out in February-September 2018.  In total, 64\,hours of telescope time were used for searching for CO(3--2) emission. The data were collected with the VEGAS spectrometer using the wide bandwidth modes 1 and 3.  The observations were done using the sub-reflector nodding (``SubBeamNod") observing mode with a six-second switching period between the two Q-band beams that are separated by 58\,arcsec on the sky.  At the central observed frequency of 42.5\,GHz, the full width at half maximum (FWHM) resolution of the GBT beam is about 17\,arcsec. The bright nearby quasar 3C279 was used for pointing and focus, while 3C286 was used for absolute flux calibration \citep{perley17}.  After correcting for atmosphere, pointing drifts, and calibration, the uncertainty on the final flux density scale is approximately 15\%.

\subsection{HST data}

We reprocessed all available \textit{Hubble} Space Telescope (HST) and {\it Spitzer} data in archive in the A1689 field as a part of the Complete Hubble Archive for Galaxy Evolution (CHArGE; Brammer, in prep), which aims to perform uniform processing and analysis of all archival HST and \emph{Spitzer} data taken away from the Galactic midplane \citep{fujimoto2022}. 
We aligned all of the HST exposures to sources in the \emph{Gaia} DR2 catalog \citep{gaia2018} and created final mosaics in a common pixel frame with 50\,mas and 100\,mas pixels for the Advanced Camera for Surveys (ACS) Wide Field Channel (WFC) and Wide Field Camera 3 (WFC3) IR filters, respectively.  
We aligned the individual \emph{Spitzer} exposures to the same astrometric frame as the HST frame and generated final drizzled IRAC mosaics with $0\farcs5$ pixels.  
Further details of the HST (\emph{Spitzer}) image processing with the \texttt{grizli} (\texttt{golfir}) software is presented in \citet{kokorev22}.

\section{Results}
\label{sect:res}

\begin{figure}
    \includegraphics[width=0.538\columnwidth]{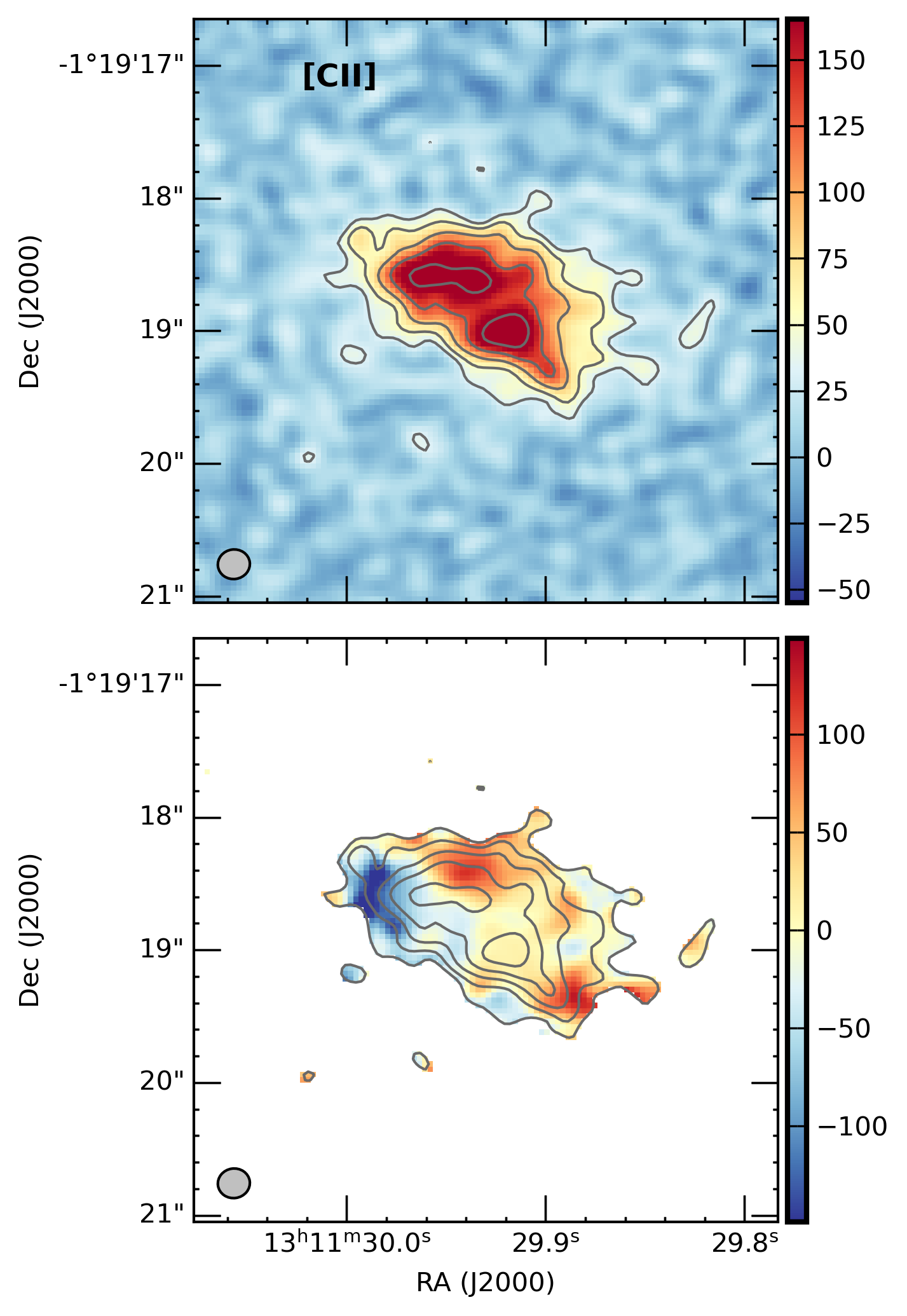}\includegraphics[width=0.462\columnwidth]{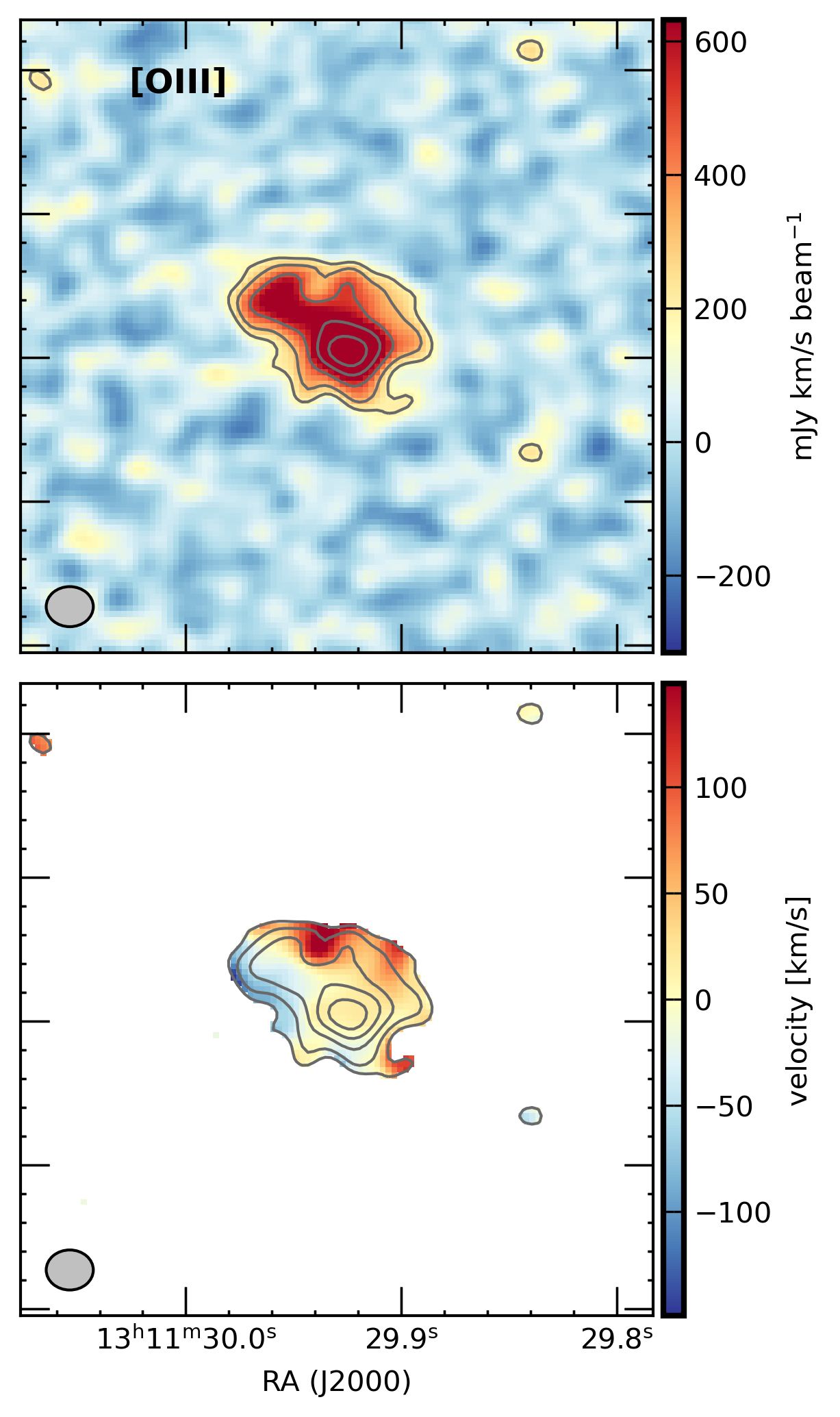}
    \caption{Velocity-integrated maps of \cii (left) and \oiii (right) of A1689-zD1 with the moment-0 (top) and moment-1 (bottom), derived from the velocity ranges $v_{\rm [CII]} = -215-270$\,km\,s$^{-1}$ and $v_{\rm [OIII]} = -250 - 270$\,km\,s$^{-1}$, respectively. 
    The contours represent the $3,5,8,12$, and $17\sigma$ levels, where $\sigma_{\rm [CII]} = 11.0$\,mJy\,km\,s$^{-1}$\,beam$^{-1}$ and $\sigma_{\rm [OIII]} = 63.2$\,mJy\,km\,s$^{-1}$\,beam$^{-1}$. The moment-1 map is masked at $3\sigma$ of the moment-0 map. 
    }
    \label{fig:momentmaps}
\end{figure}

\subsection{\cii and \oiii}

\begin{figure*}
\includegraphics[width=\textwidth]{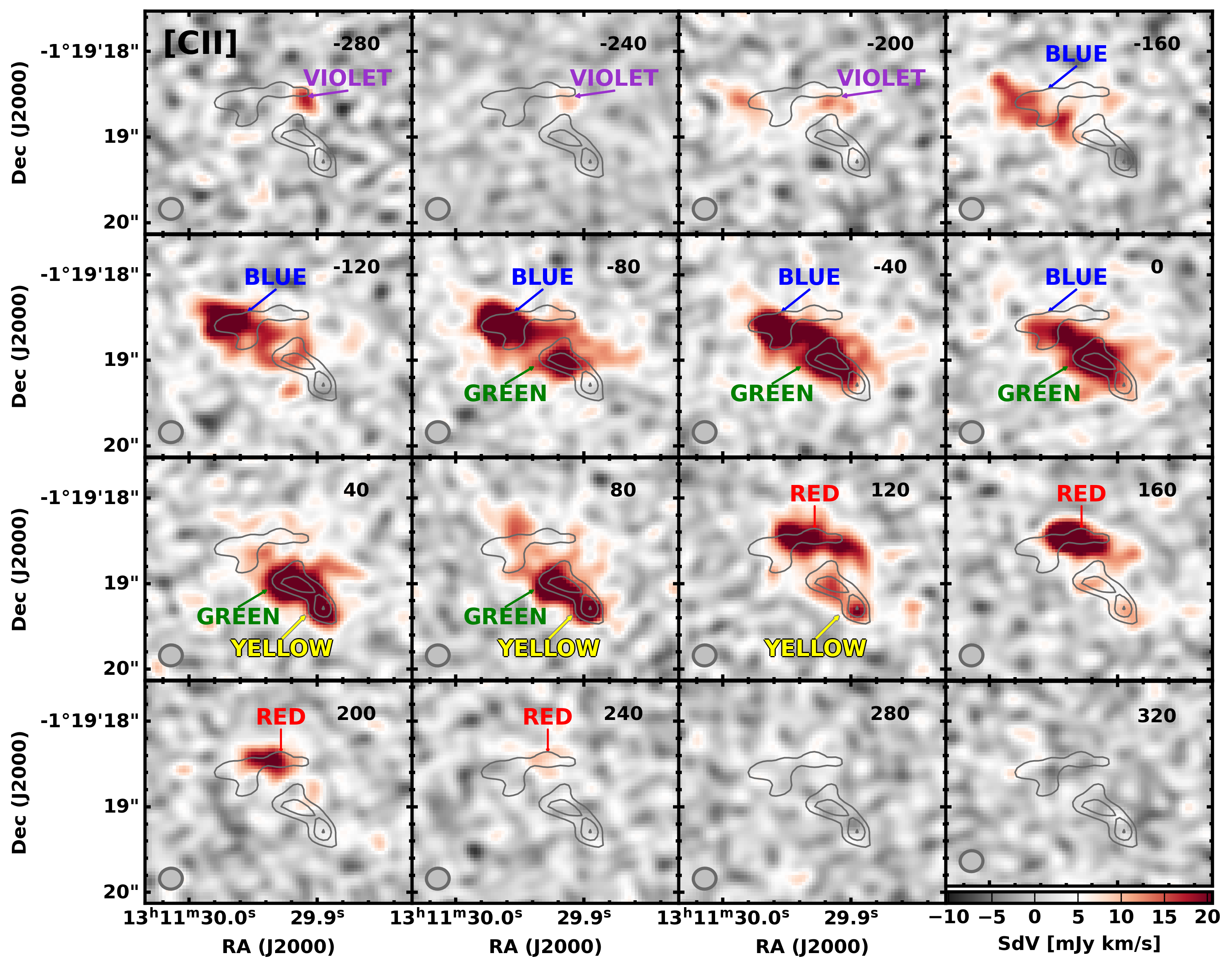}
\caption[]{Mosaic of the \cii binned by velocity in 40\,km\,s$^{-1}$ intervals. The value in the upper right corner of each sub-image indicates the middle of the velocity relative to $z=7.133$.  The gray contours show the band 6 continuum, where the levels correspond to $4, 6, 8\sigma$.  The five components of the analysis (Sect.~\ref{subsect:linecontdecomp}) are indicated by name. 
The color bar below the last panel shows the level corresponding to the velocity integrated flux, $S{\rm d}V$, for each map.
\label{fig:ciimosaic}
}
\end{figure*}

\begin{figure*}
\includegraphics[width=\textwidth]{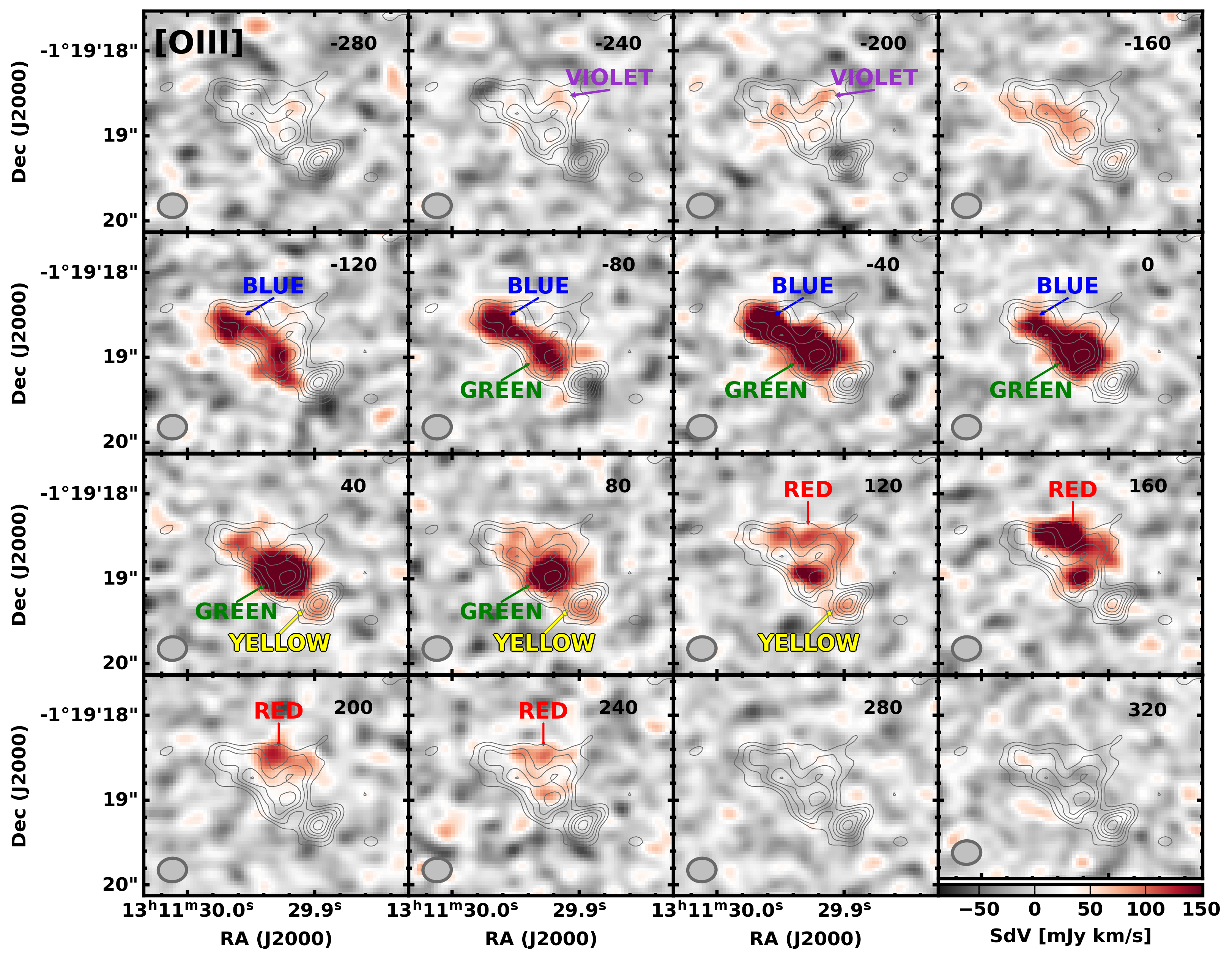}
\caption[]{Mosaic of the \oiii observations. The upper right corner shows the central
velocity of each 40\,km\,s$^{-1}$ channel relative to $z=7.133$.  The gray contours show the band 8 continuum, where the levels correspond to $4, 6, 8, 10, 12$, and $14\sigma$.  The five components of the analysis (Sect.~\ref{subsect:linecontdecomp}) are indicated by name.
The color bar below the last panel shows the level corresponding to the velocity integrated flux, $S{\rm d}V$, for each map.
\label{fig:oiiimosaic}
}
\end{figure*}

Emission lines are detected in both ALMA data sets consistent with \cofe and \oee. Both lines show
a complex line profile as well as complex surface brightness distribution.  
Similarly, the continuum emission shows a complex structure consistent
with the integrated emission line distribution.  
In Fig.~\ref{fig:momentmaps}, we show the moment-0 and moment-1 maps, while in Figs.~\ref{fig:ciimosaic} and \ref{fig:oiiimosaic}, we show the line emission per 40\,km\,s$^{-1}$ channel for both lines.  The integrated line intensity along with continuum flux densities are given in Table~\ref{tab:fluxresults}. The band 6 and band 8 continuum results are consistent with the photometric results from \citet{knudsen17} and \citet{inoue20}.

\begin{table}
    \centering
    \caption{Integrated line intensity and continuum flux density.} 
    \begin{tabular}{l>{\hspace{-5ex}}rr>{\hspace{-1ex}}l@{}}
    \hline
    \hline
       Line  &  $I_{\rm line}$ (Jy\,km\,s$^{-1}$)   & $L_{\rm line}$ \\
    \hline
      \cii   & $ 4.4\pm 0.1$ & $ (5.8\pm0.1)\times 10^{8}$&L$_\odot$\\
      \oiii & $5.3\pm 0.1$ & $(1.3\pm0.02)\times10^9$&L$_\odot$\\
      CO(3--2) &  $< $0.034 & $< 6.2\times 10^{8}$&K\,km\,s$^{-1}$\,pc$^2$ \\
      CO(6--5)\tablefootmark{a}  &  $<0.68$ & $<3.1\times10^9$&K\,km\,s$^{-1}$\,pc$^2$ \\ 
      CO(7--6)\tablefootmark{a}  &  $<0.64$ & $<2.1\times10^9$&K\,km\,s$^{-1}$\,pc$^2$ \\
      \ci\tablefootmark{a}  & $<0.64 $ & $<3.8\times10^7$&L$_\odot$\\
    \hline\\[-6pt]
    Continuum  & $S_{\nu}$ (mJy) & \\
     \hline
      417\,GHz  & $1.5\pm0.13$ & \\
      233\,GHz  & $0.49\pm0.032$ & \\ 
      91\,GHz$^a$ & $<47$ & \\
    \hline
    \end{tabular}
    \tablefoot{ 
    The line luminosity is corrected for gravitational lensing magnification.\\
    \tablefoottext{a}{Assuming the source is extended over two beams in the band 3 data.} 
    } 
    \label{tab:fluxresults}
\end{table}

\subsection{Redshift}

The \cofe and \oee lines suggest central redshifts of 7.1329 and 7.1334, respectively, based on a single Gaussian profile fit. Taking the mean of these as the systemic redshift and the difference as the uncertainty, we find A1689-zD1 is at \(z=7.1332\pm0.0005\). This is significantly ($\sim13\,000$\,km\,s$^{-1}$) offset from the redshift derived from the NIR spectrum of the Lyman-break in \citet{watson15}, which was \(7.5\pm0.2\). 
This is similar to the 11\,000km\,s$^{-1}$ difference seen in redshift determination
using optical/NIR spectroscopy and ALMA spectroscopy for
MACSJ1149-JD at \(z = 9.11\) in \oiii\ emission \citep{hashimoto18} and the photometric redshift of 9.51 \citep{hoag18}, as well as for GNz11 of $z_{\rm grism} = 11.1$ versus $z_{\rm spec} = 10.6$ confirmed with the JWST \citep{oesch16,bunker23}.
As bright Lyman-alpha emission or other  UV emission lines have not
been detected in the spectra, the optical redshift has been derived based on
the continuum break. However, given the potential large column of HI along
the line of sight as well as the relative faintness of these distant
galaxies provides a challenge for obtaining redshifts. A detailed analysis of
this will be provided in Watson et al. (in~prep.).

\subsection{Molecular gas}

\begin{figure}
\includegraphics[width=\columnwidth]{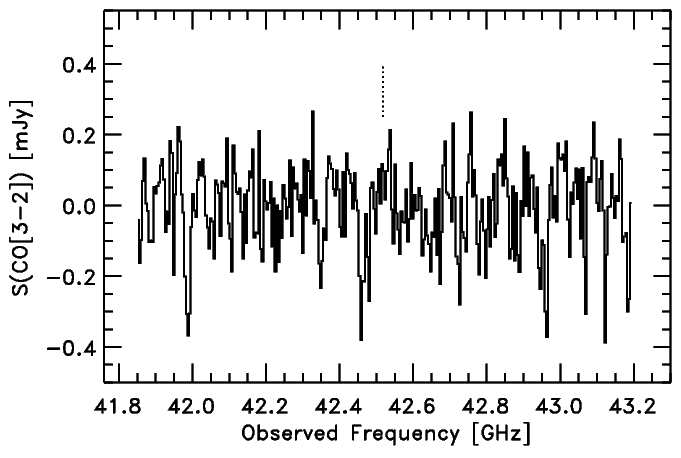}
\caption[]{Non-detection of CO(3--2) emission measured at the GBT smoothed to a channel resolution of about 30\,km\,s$^{-1}$.  The expected line location based on the \cii and \oiii redshift of $z=7.133$ is shown by the vertical dotted line.}
\label{fig:gbtco32}
\end{figure}

Figure~\ref{fig:gbtco32} shows the result of the GBT observations.  The 
CO(3-2)
line was not detected, and the $3\sigma$ limit is given in Table~\ref{tab:fluxresults} 
assuming a linewidth of 300\,km\,s$^{-1}$.
To mitigate residual baseline structures  high-pass filtering methods were applied to the individual SubBeamNod data scans.  These filtering methods would remove any underlying broad features ($>1000$\,km\,s$^{-1}$), while preserving emission for the expected line width of about 300\,km\,s$^{-1}$ and, thus,  resulting in a flat spectrum (Fig. \ref{fig:gbtco32}).  Although the line was not detected, if we were to smooth the data further, there is a weak, nearly $2\sigma$ bump (with a width of about \(200\pm100\)\,km\,s$^{-1}$ at $z=7.13$ when fit by a Gaussian) at the expected frequency.  Hundreds of hours of additional GBT time would be required to test whether this ''bump'' is real.

\begin{figure}
\includegraphics[width=\columnwidth]{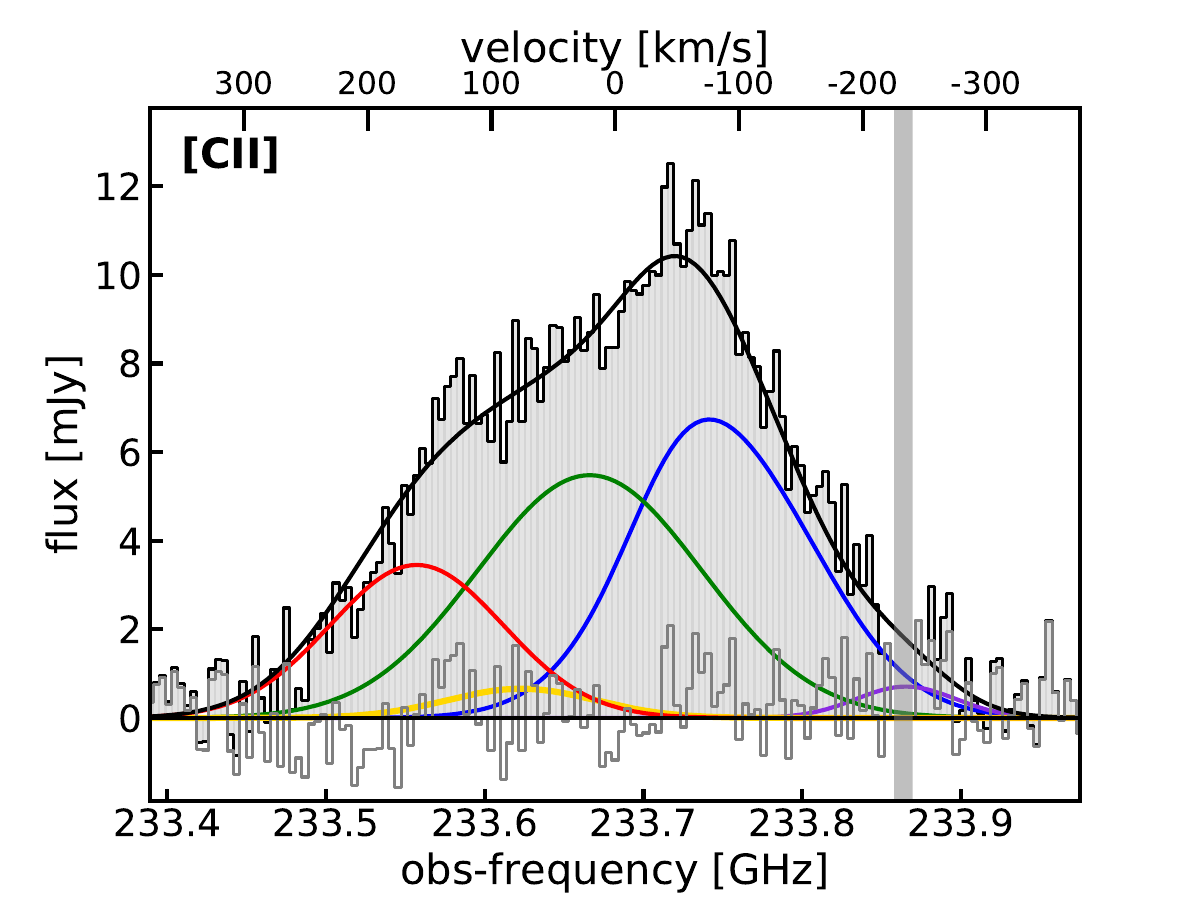}
\includegraphics[width=\columnwidth]{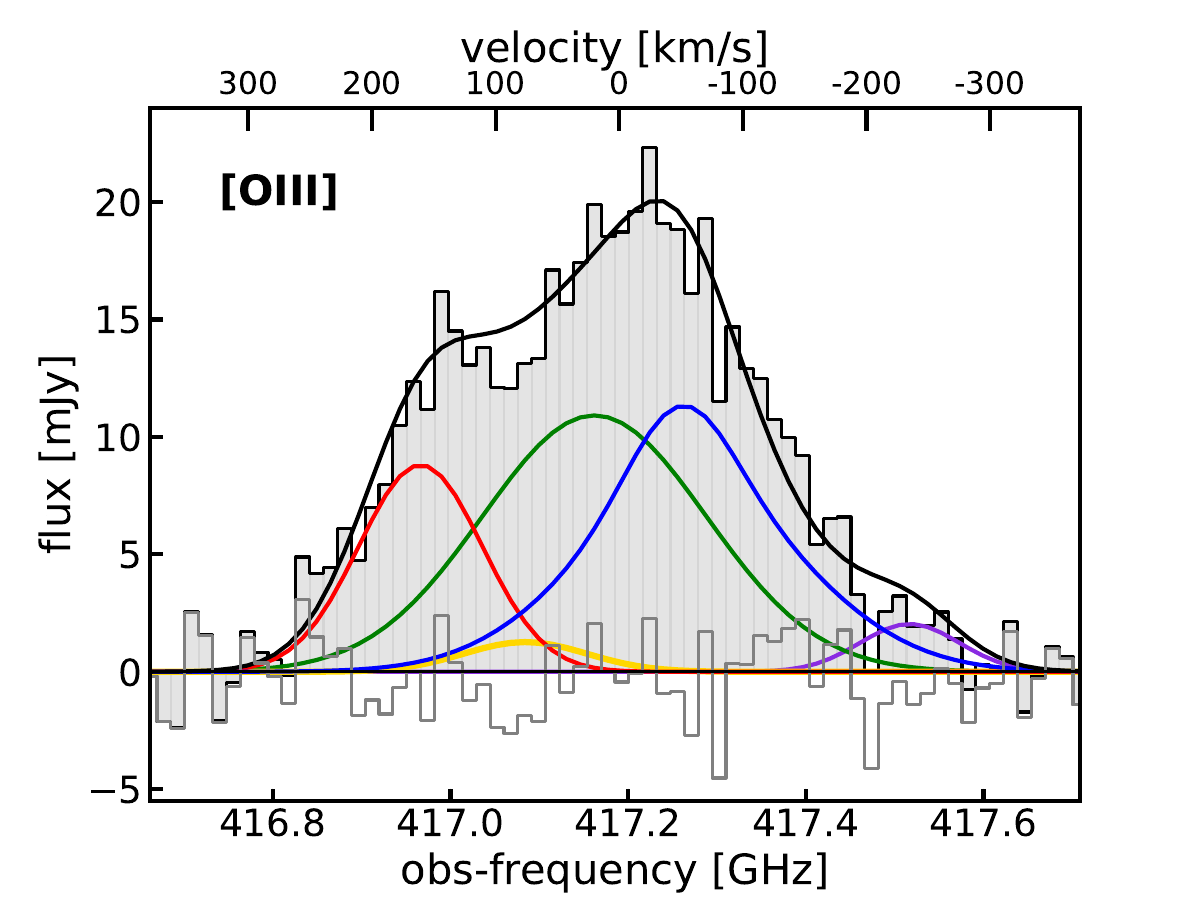}
\caption[]{\cii ({\it top}) and \oiii ({\it bottom}) spectra (continuum-subtracted) overlaid with the resulting models for each component (color coded according to their component names). The black line shows the sum of the models. The residuals after subtracting the model results are shown as a thin gray histogram.  
\label{fig:cii_spec_model}}
\end{figure}

The CO(6--5), CO(7--6), and the \ci\!(1-0)\,492 GHz lines were not detected in the ALMA band 3 data. Assuming a linewidth of 300\,km\,s$^{-1}$ and an area similar to the \oiii detection, we estimated $3\sigma$ upper limits for all three transitions. These are given in Table~\ref{tab:fluxresults}. 
We note that detection of moderately high-$J$ CO transitions is a challenge of $z>6$ galaxies \citep[e.g.,][]{hashimoto23}.

\subsection{Gravitational lensing}

We used a parametric model of the Abell~1689 mass distribution to estimate the magnification and distortions at the location of A1689-zD1, as well as across the source. Our reference mass model is constructed with the \texttt{Lenstool} software \citep{jullo07} and described in \citet{limousin07}. 

At the measured spectroscopic redshift of $z = 7.133$ this model predicts a single image at this location with a total magnification of 9.6, involving an angular magnification of $\sim$6$\times$, along a specific direction at ${\rm PA}=85$ degrees (roughly east-to-west), and $1.6\times$ perpendicular to it.
The amplification factor and shear directions do not vary by more than 10\% and $5\degr$, respectively, across the 1.5\arcsec extent of the source (Table~\ref{tab:comp_mu}). This is due to the fact that there is no significant cluster substructure or cluster member identified in the vicinity of A1689-zD1, which could locally affect the mass distribution and magnification.

To estimate the errors on the magnification factor, we used the best-fit mass model and the range of Markov chain Monte Carlo samples from the optimisation performed in \texttt{Lenstool} by \citet{limousin07}. We obtain a statistical relative error of 19\% (Table~\ref{tab:comp_mu}). In addition, we make use of additional existing mass models of the same cluster (e.g.,\ \citealt{Xue}) where a different value is used for the relative contribution of the cluster-scale and galaxy-scale distribution of the mass model, to estimate the systematic uncertainty on the average magnification of the source. We estimate a systematic relative error of 10\%.

The angular magnification implies an angular resolution in the \cii map of
$\sim0.037''\times0.15''$, which correspond to \(0.2\,\mathrm{kpc}\times 0.78\,\)kpc. 
Correcting the sensitivity for the magnification, this implies a spectroscopic
rms of $\sim20\,\mu$Jy/beam per 5\,km\,s$^{-1}$ channel and a continuum sensitivity of
0.67\,$\mu$Jy/beam for the band 6 data, making this some of the deepest data
obtained for a single high-$z$ galaxy with ALMA.  

For the \oiii data, the magnification-corrected resolution is $\sim 0.051''\times 0.18''$ corresponding to \(0.27\,\mathrm{kpc}\times0.95\,\)kpc.  The spectroscopic sensitivity corresponds to 42--47\,$\mu$Jy/beam in a $\sim11$\,km\,s$^{-1}$ channel and the continuum sensitivity is $3.1\,\mu$Jy/beam.

\begin{table}
    \centering
    \caption{Individual magnification factors for the different components.}
    \label{tab:comp_mu}
    \begin{tabular}{lc}
\hline
\hline
        Component & $\mu$   \\
                  &     \\
\hline
        \textit{Violet} & 9.44$\pm$1.79  \\
        \textit{Blue1} & 9.44$\pm$1.79     \\
        \textit{Blue2} & 9.38$\pm$1.78  \\
        \textit{Green} & 9.62$\pm$1.83  \\
        \textit{Yellow} & 9.84$\pm$1.87  \\
        \textit{Red1} & 9.34$\pm$1.77 \\
        \textit{Red2} & 9.48$\pm$1.80 \\
\hline
    \end{tabular}
\end{table}

\subsection{Decomposition into galaxy components}
\label{subsect:linecontdecomp}

When stepping through the continuum-subtracted ALMA line datacubes, both \cii (Fig.~\ref{fig:ciimosaic}) and \oiii (Fig.~\ref{fig:oiiimosaic}) show emission peaking at different spatial locations in the source as a function of velocity. This complex structure is dominated by three main spatial components with a maximum flux at $\sim-80$ ("blue"), $\sim0$ ("green"), and $\sim+160$ km\,s$^{-1}$ ("red") and present in both emission lines. This multi-component structure  was already evident from the complexity seen in previous continuum results \citep{knudsen17}. However, emission lines at a high S/N now provide us with a better possibility for disentangling this scenario. 

First, as  made clear from the relative spatial locations of the main components, as well as the moment-1 map derived from the cube (Fig.~\ref{fig:momentmaps}), they do not follow a regular structure (as seen, e.g., in regular rotating disk galaxies). Therefore, we identified and model each component from the datacubes using an iterative process: (1) we produce a narrowband image from the line data comprising the main channels of emission in each component; (2) we fit this image with a 2D elliptical Gaussian model for the line flux distribution (including the central location, total flux, full width at half maximum (FWHM) along major and minor axes, and PA of the major axis); (3) we extracted the corresponding spectrum of this component and fit line parameters with a 1D Gaussian (central velocity, $v_0$, velocity dispersion, $\sigma_0$); and (4) we subtracted a 3D model of line emission for this component combining the spatial and spectral model.

Due to the proximity of some components (both spatially and in velocity space), this fit is performed three times iteratively,  subtracting all other components from the datacube before a new fit. In addition to the three main components mentioned above, two fainter components peaking at $\sim +80$ ("yellow") and $\sim-220$ ("violet") km\,s$^{-1}$ were added, and more complex spatial distributions of the flux were used when necessary from the residuals. A few extended components also revealed a spatial variation of the line peak, which we fit using an empirical velocity gradient model (arctan model with a truncation velocity, $v_t$, and a truncation radius, $r_t$). We summarize the decomposition into five components below and we provide the full model parameters in Tables~\ref{tab:CIIcomponents} and \ref{tab:OIIIcomponents}. We note that we originally identified the components from the \cii data, but we recovered the same structure in the \oiii data, where we followed similar narrowband images and spatial distributions for the fit of each component.

{Violet}:   The faintest of the five components. It was fit with a single non-rotating component and for \cii, the fit was made challenging by empty channels between the spectral windows.   

{Blue}:  Modeled with two elliptical Gaussian components. One is extended with rotation (denoted "blue1") and one compact with only weak rotation ("blue2"). Their central positions are close but not identical. 

{Green}: Broad component modeled with a single extended source, which has weak rotation.       

{Yellow}: Single compact component with no measurable rotation. 

{Red}:  Combination of two individual elliptical Gaussian profiles with weak rotation, extended in the WSW-ENE direction.     

Combining these five components,
we recovered 95 and 97\% of the total emission measured in the \cii and \oiii line data, respectively, (Fig.~\ref{fig:cii_spec_model}) in the inner region (not the extended region studied in \citealt{akins22}). Some extra emission, somewhat diffuse, appears to be present but cannot easily be modeled as individual components. It is worth noting that the choice of decomposition (number of components and their  parametrization), while somewhat arbitrary, is driven by the complexity and the high signal-to-noise ratio (S/N) of the data. Correcting the data for gravitational magnification, the extracted components
are found to have sizes in the range 0.1--1\,kpc.

\begin{table*}
    \centering
    \caption{\cii model results. }
    \label{tab:CIIcomponents}
    \begin{tabular}{@{}lcccccccccccc@{}}
\hline
\hline
       Component &$x_c$ & $y_c$&FWHM$_1$ & FWHM$_2$ & $\theta$ & Flux & $v_0$ & $\sigma_0$ & $v_t$ & $r_t$ & $i$ & $\theta_v$\\
        & ["] & ["] & ["] & ["] & [deg] & [Jy\,km\,s$^{-1}$] & [km\,s$^{-1}$] & [km\,s$^{-1}$] & [km\,s$^{-1}$] & ["] & [deg] & [deg] \\
\hline
\textit{Violet} & -0.4 & 0.2 & $<$0.13 & $<$0.31 & [0.0] & 0.06$\pm$0.01 & -235 & 40 & 10 & 0.5 & [30] & 180\\
\textit{Blue1} & 0.2 & 0.1 & 0.61 & 1.32 & 80.8 & 0.85$\pm$0.02 & -90 & 75 & 60 & 2.4 & [30] & -20\\
\textit{Blue2} & 0.4 & 0.1 & $<$0.14 & $<$0.34 & [0.0] & 0.10$\pm$0.01 & -50 & 40 & 10 & 0.5 & [30] & 180\\
\textit{Green} & -0.3 & -0.3 & 0.38 & 0.57 & 96.3 & 0.55$\pm$0.02 & 20 & 90 & 40 & 2.8 & [30] & 200\\
\textit{Yellow} & -0.6 & -0.6 & $<$0.08 & $<$0.24 & [0.0] & 0.04$\pm$0.01 & 76 & 60 & 0 & 0.1 & [30] & -35\\
\textit{Red1} & 0.1 & 0.3 & 0.23 & 0.60 & 75.4 & 0.42$\pm$0.02 & 160 & 60 & 5 & 0.5 & [30] & -20\\
\textit{Red2} & -0.6 & 0.2 & 0.45 & 0.45 & 0.0 & 0.13$\pm$0.02 & 160 & 60 & 5 & 0.5 & [30] & -20\\
\hline
    \end{tabular}
\tablefoot{
Central position is provided from the location  $\alpha=$197.8747351
$\delta=-$1.3218751. The FWHM values are corrected for the beam, but not for
the gravitational magnification. Values in square brackets are fixed
parameters. Correcting the data for gravitational magnification, the extracted components
are found to have sizes in the range 0.1--1\,kpc. 
}
\end{table*}

\begin{table*}
    \centering
    \caption{\oiii model results.}
    \label{tab:OIIIcomponents}
    \begin{tabular}{@{}lcccccccccccc@{}}
\hline
\hline
       Component &$x_c$ & $y_c$&FWHM$_1$ & FWHM$_2$ & $\theta$ & Flux & $v_0$ & $\sigma_0$ & $v_t$ & $r_t$ & $i$ & $\theta_v$\\
        & ["] & ["] & ["] & ["] & [deg] & [Jy\,km\,s$^{-1}$] & [km\,s$^{-1}$] & [km\,s$^{-1}$] & [km\,s$^{-1}$] & ["] & [deg] & [deg] \\
\hline
\textit{Violet} & -0.3 & 0.1 & 0.23 & 0.23 & 0.0 & 0.20$\pm$0.01 & -235 & 40 & 10 & 0.5 & [30] & 180\\
\textit{Blue1} & 0.2 & 0.0 & 0.46 & 0.84 & 57.4 & 1.66$\pm$0.03 & -60 & 90 & 60 & 2.4 & [30] & -20\\
\textit{Blue2} & 0.4 & 0.2 & $<$0.14 & $<$0.39 & [0.0] & 0.28$\pm$0.01 & -50 & 40 & 10 & 0.5 & [30] & 180\\
\textit{Green} & -0.2 & -0.2 & 0.31 & 0.41 & 83.4 & 2.36$\pm$0.02 & 20 & 90 & 40 & 2.8 & [30] & 200\\
\textit{Yellow} & -0.6 & -0.6 & $<$0.1 & $<$0.39 & [0.0] & 0.20$\pm$0.01 & 76 & 50 & 0 & 0.1 & [30] & -35\\
\textit{Red1} & 0.1 & 0.3 & 0.14 & 0.54 & 81.0 & 0.78$\pm$0.03 & 160 & 50 & 5 & 0.5 & [30] & -20\\
\textit{Red2} & -0.5 & 0.1 & 0.32 & 0.32 & 0.0 & 0.38$\pm$0.01 & 160 & 50 & 5 & 0.5 & [30] & -20\\
\hline 
    \end{tabular}
\tablefoot{Central position is provided from the location  $\alpha=$197.8747351 $\delta=-$1.3218751. The FWHM values are corrected for the beam, but not for the gravitational magnification. Values in square brackets are fixed parameters.}
\end{table*}

\begin{figure*}
\begin{center}
\includegraphics[width=0.80\textwidth]{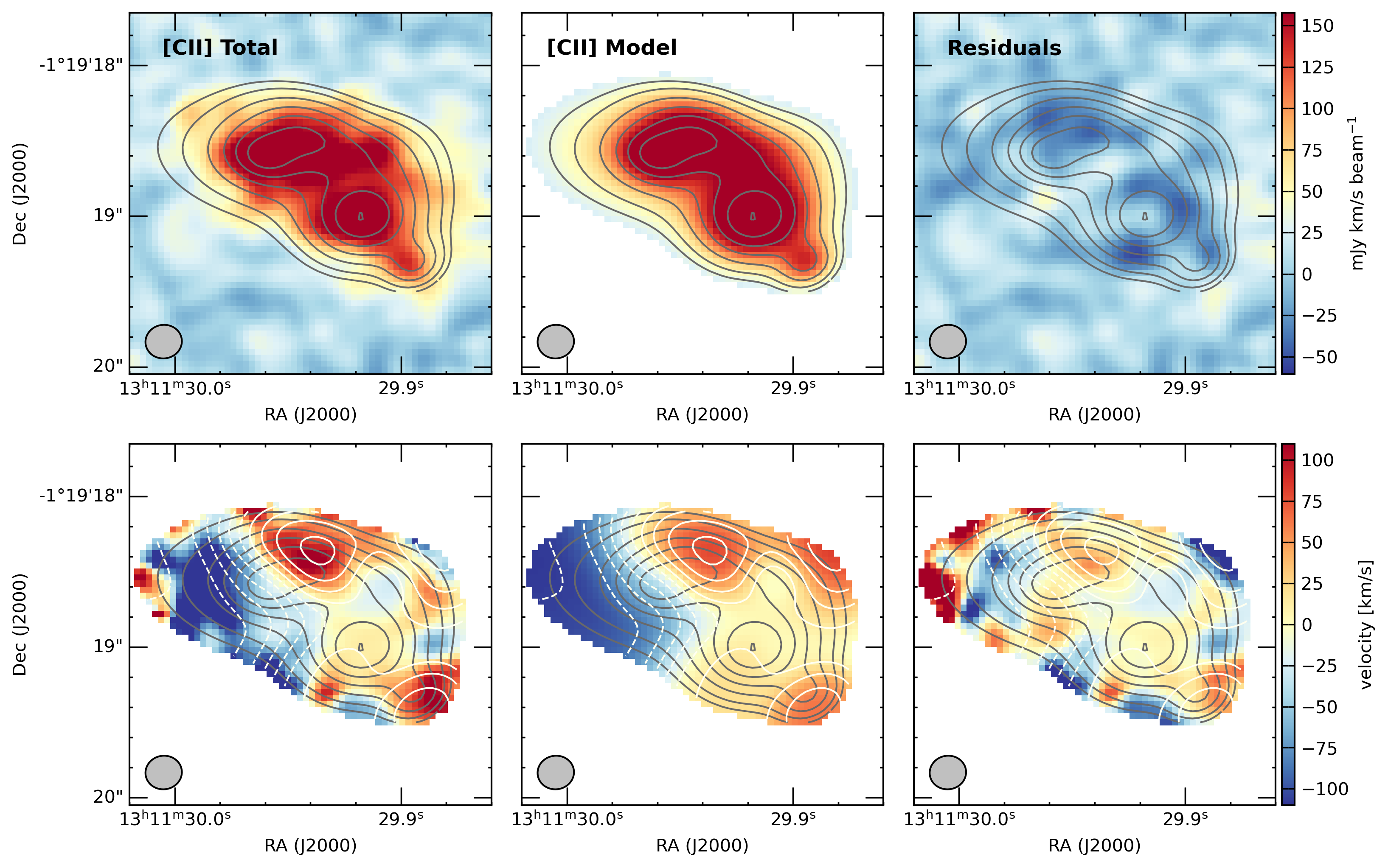}
\includegraphics[width=0.80\textwidth]{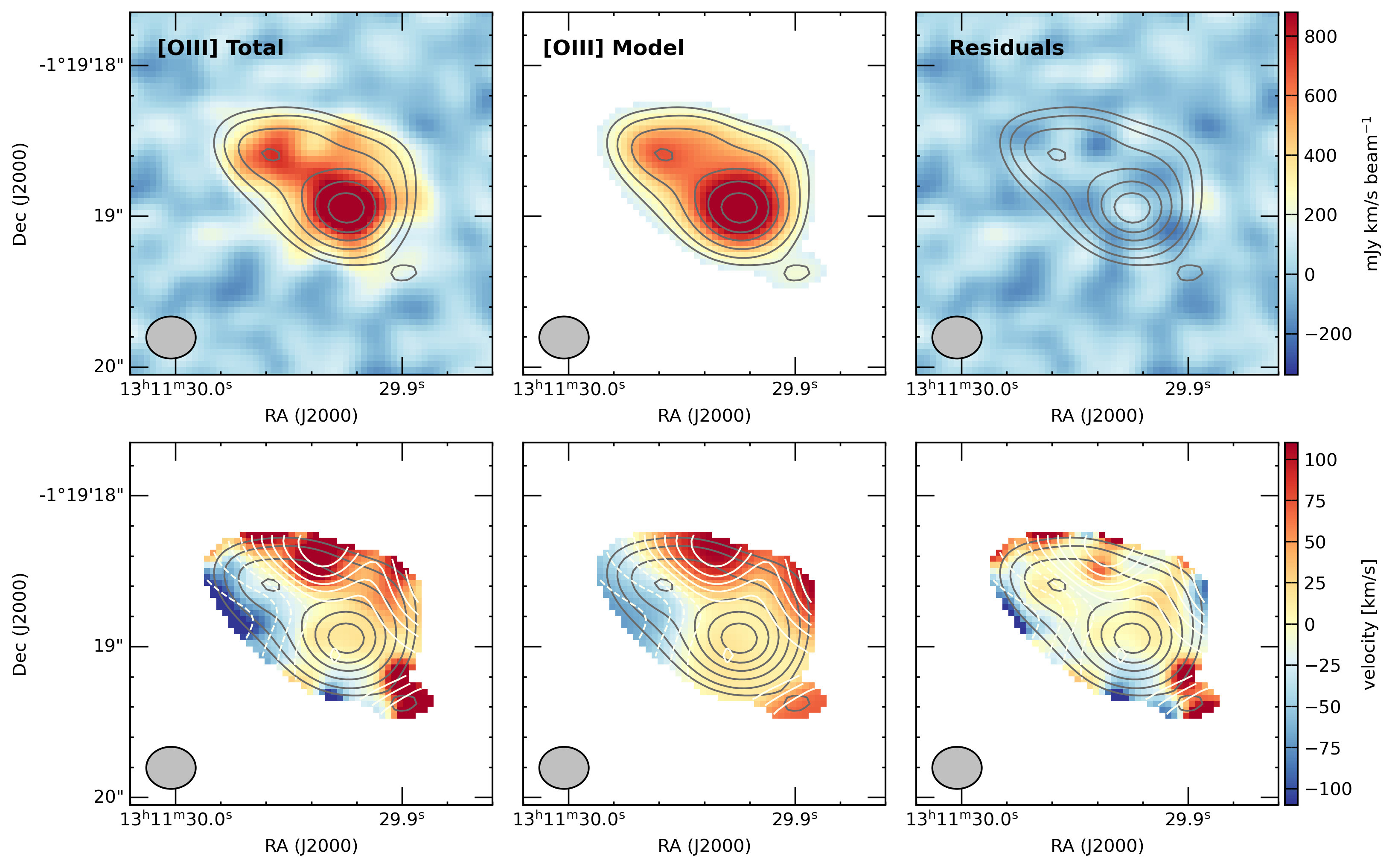}
\end{center}
\caption[]{Overall moment-0 maps (first and third rows) and moment-1 maps (second and fourth rows) for A1689-zD1 for \cii and \oiii\ for the data (left), the model (middle), and  the  residuals (right). 
Contours show the model with the gray showing the moment-0 (3, 4, 5, 7, 10, 13, 16, and 19$\sigma$, where the rms is measured in the data regions excluding the source), and the white contours showing the moment-1 in steps of 20\,km\,s$^{-1}$. The applied mask is determined from the model. 
We note that this does not include the extended "halo" component presented in \citet{akins22}. 
\label{fig:cii_model_res}}
\end{figure*}

For comparison of the derived model to the data, we show the moment-0 and moment-1 maps of both the observations and the model in Fig.~\ref{fig:cii_model_res} for both \cii and \oiii. The residual maps indicate that the models effectively reproduce the observations. We note that for \cii there could be a small offset and that this could be due to the \cii emission having an additional extended component \citep[e.g.,][]{akins22}, which adds to the complexity of the source.

\begin{figure}
\includegraphics[width=\columnwidth]{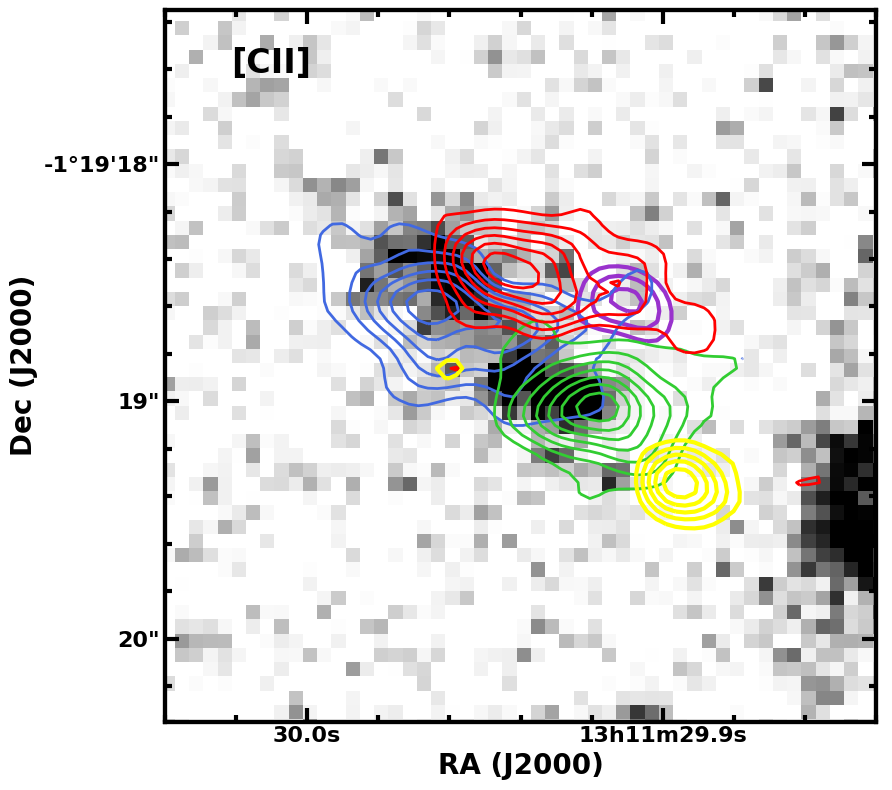}
\includegraphics[width=\columnwidth]{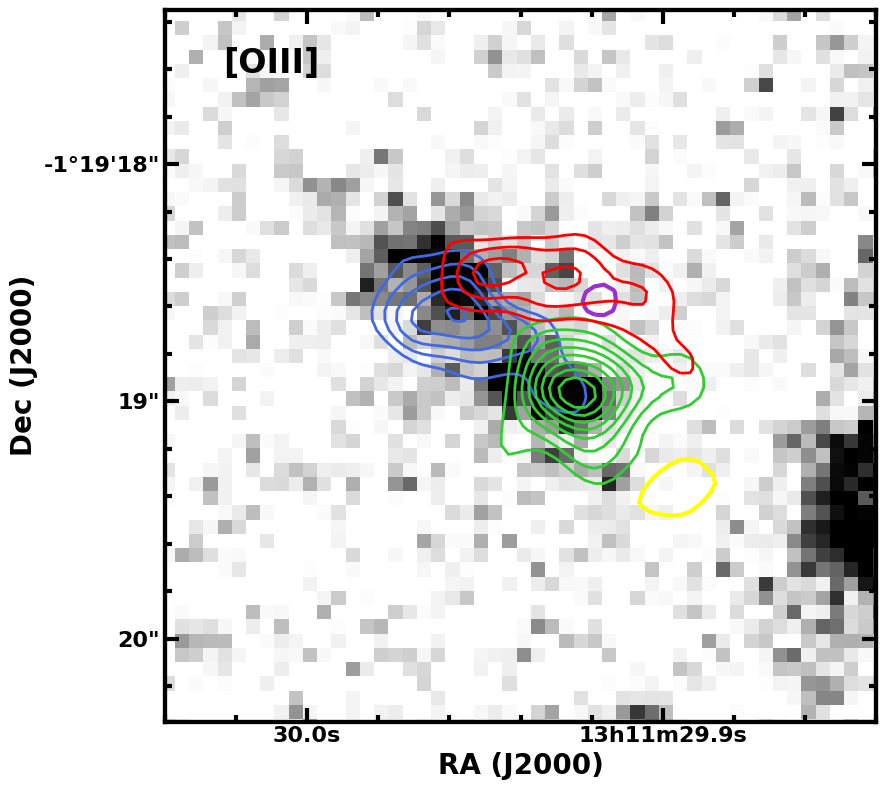}
\caption[]{Contours of the individual components of the \cofe (\textit{top}) and \oee (\textit{bottom}). Overlaid is the combined HST image obtained with F125W,
F140W, and F160W filters. The contours are color-coded according to their component name; contours are in steps of 2$\sigma$ starting at 4$\sigma$ (see Fig.~\ref{fig:components_imm01} for the rms). 
\label{fig:cii_HST}}
\end{figure}

\begin{figure*}
\begin{center}
\includegraphics[width=\textwidth]{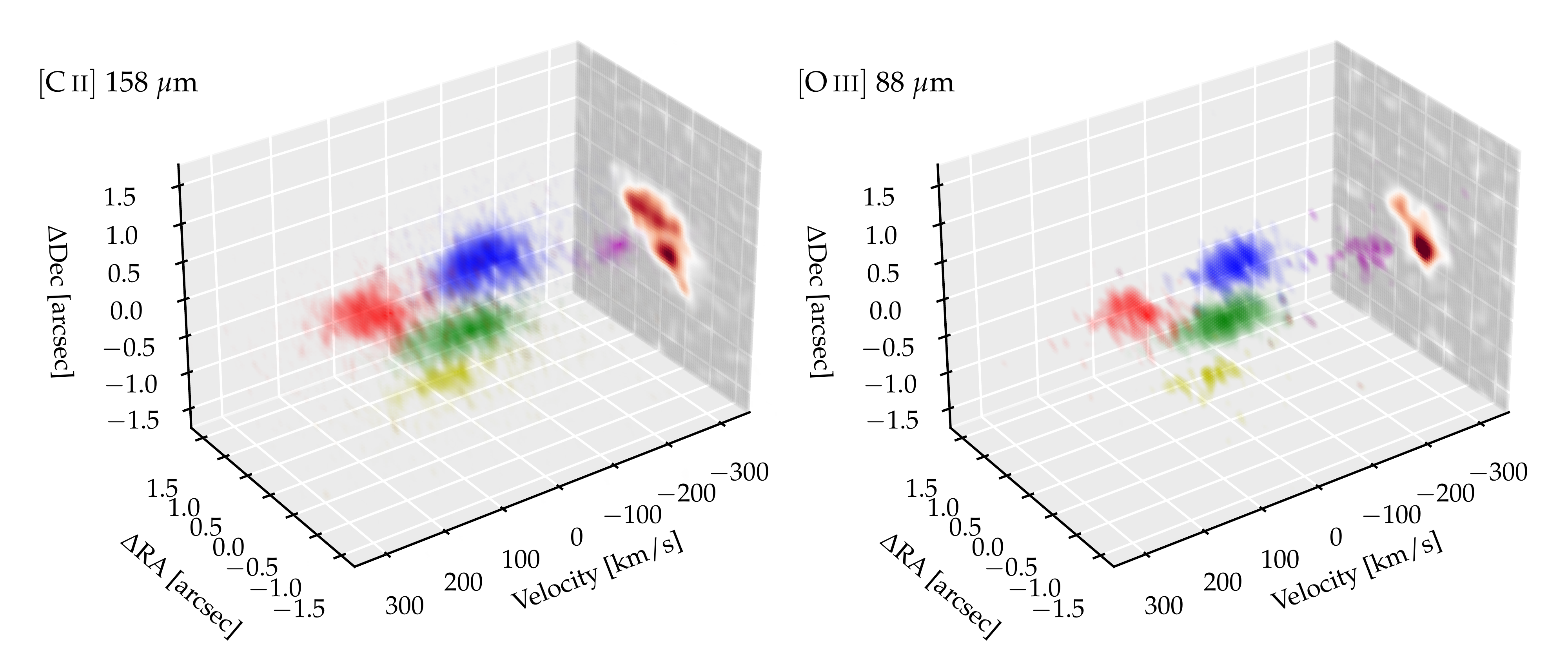}
\end{center}
\caption[]{Decomposed components of the observations of A1689-zD1 shown in 3D with color-coding according to component name. The integrated moment-0 maps is shown at the end of the plot. 
\label{fig:3d}}
\end{figure*}

In Fig.~\ref{fig:cii_HST} we show the contours of the individual components from \cii and \oiii overlaid on the HST images.  
The relative astrometry of the HST and ALMA images should be correct to \(0.^{\prime\prime}03\) allowing us to make a robust comparison of the galaxy's UV and the ALMA results.
As the HST data show no significant systematic color change across the galaxy, we used a stacked image combining the data from observations using the F125W, F140W, and F160W HST filters to increase the S/N of the rest-frame UV data. For further illustration of the data, we show the decomposed components in the 3D in Fig.~\ref{fig:3d}, using an approach to showing 3D data similar to that of \citet{spilker22} and showing the emission of each component above S/N of 3.

The clearest difference between the UV and FIR data is the number of components, with two components seen in the UV and the five components of the \cii and \oiii data.  Both the violet and yellow components have no obvious UV-counterpart. Given the apparent offsets between the green, blue, and red components, we cannot unambiguously associate the components with the UV-components. One possible interpretation is that green is associated with the SW UV-component and blue is associated with the NE UV-component. 
In contrast, the 
violet and yellow components do not show HST counterparts, while dust continuum emission is detected from these components.
This suggests that completely dust-obscured regions exist in this system, consistent with the highly ($\simeq80-90$\%) obscured SFR fraction of A1689-zD1 \citep{bakx21,akins22}. This is also reminiscent of the recent identification of completely dust-obscured HST-dark galaxies at $z>7$ \cite{fudamoto21}. 
We further discuss the possible origins of these components in Sect.~\ref{sect:disc} below.

\begin{table*}
    \caption{
   Dust continuum decomposition.
    }
    \label{tab:dustcomponents}
    \begin{tabular}{@{}lccccccccc@{}}
\hline
\hline
Comp. & $S_{\rm 88 \mu m, ap}$ & $S_{\rm 158 \mu m, ap}$ & $S_{\rm 88 \mu m, mod}$ & $S_{\rm 158 \mu m, mod}$ & $T_{\rm dust, ap}$ & $L_{\rm IR, ap}$ & $T_{\rm dust, mod}$ & $L_{\rm IR, mod}$ & $L_{\rm IR, 41K}$  \\
& [mJy] & [mJy]  & [mJy]   & [mJy]  & [K] & [$10^{10}L_{\odot}$] & [K]  & [$10^{10}L_{\odot}$] & [$10^{10}L_{\odot}$]  \\[3pt] \hline
\textit{Violet} & $0.15\pm0.02$ & $0.03\pm0.01$ & $0.20\pm0.04$ &
\ \ $0.01\pm0.01$\tablefootmark{a}  & 68$_{-23}^{+46}$ & 4.9$_{-3.3}^{+16}$  &  \nodata & \nodata & 1.1$_{-0.6}^{+1.8}$ \\[3pt]
\textit{Blue}   & $0.30\pm0.01$ & $0.09\pm0.02$  & $0.57\pm0.10$ & $0.16\pm0.04$  & 47$_{-7}^{+8}$ & 3.7$_{-1.0}^{+1.5}$ &  53$_{-13}^{+17}$ & 8.9$_{-4.9}^{+11}$ & 5.0$_{-2.1}^{+6.2}$ \\[3pt]
\textit{Green}  & $0.38\pm0.01$ & $0.10\pm0.02$  & $0.68\pm0.13$& $0.18\pm0.03$  & 56$_{-9}^{+11}$ & 7.0$_{-4.7}^{+12}$ &  56$_{-13}^{+19}$ & 13$_{-6}^{+17}$ & 5.3$_{-2.5}^{+7.8}$ \\[3pt]
\textit{Yellow} & $0.22\pm0.02$ & $0.06\pm0.01$  & $0.25\pm0.02$ & $0.07\pm0.02$ & 55$_{-12}^{+16}$ & 3.8$_{-1.1}^{+4.0}$ &  49$_{-10}^{+13}$ & 3.3$_{-1.2}^{+2.8}$ & 2.4$_{-0.5}^{+2.4}$ \\[3pt]
\textit{Red}    & $0.15\pm0.02$ & $0.04\pm0.01$  &
\ \ $0.14\pm0.08$\tablefootmark{a} & $0.06\pm0.01$ & 62$_{-21}^{+38}$ & 3.6$_{-2.2}^{+14}$ &  \nodata & \nodata & 1.5$_{-0.7}^{+2.2}$ \\[3pt]
\hline 
\end{tabular}
\tablefoot{
Infrared luminosity, $L_{\rm IR}$, and SFR estimates are corrected for
gravitational lensing.\\
\tablefoottext{a}{We extracted the photometry by using a $0\farcs2$ circular
radius aperture placed on the residual map, instead of including this
component in the model.}
}
\end{table*}

\begin{table*}
    \centering
    \caption{Integrated line intensity and line luminosity per component.}
    \label{tab:resolved_linelum}
    \setlength{\tabcolsep}{3.5pt}
    \begin{tabular}{@{}lcccccccccc@{}}
\hline
\hline
Comp- & $I^{\rm obs}_{\rm [CII]}$ & $I^{\rm mod}_{\rm [CII]}$ & $I^{\rm obs}_{\rm [OIII]}$ & $I^{\rm mod}_{\rm [OIII]}$ & $L^{\rm obs}_{\rm [CII]}$ & $L^{\rm mod}_{\rm [CII]}$ & $L^{\rm obs}_{\rm [OIII]}$ & $L^{\rm mod}_{\rm [OIII]}$ &  $L^{\rm obs}_{\rm [OIII]}$/$L^{\rm obs}_{\rm [CII]}$ &  $L^{\rm mod}_{\rm [OIII]}$/$L^{\rm mod}_{\rm [CII]}$ \\  
 onent & \multicolumn{4}{c}{\Vhrulefill~[Jy km\,s$^{-1}$]~\Vhrulefill}& \multicolumn{4}{c}{\Vhrulefill~[$10^8$\,L$_\odot$]~\Vhrulefill} &  & \\ 
\hline
\textit{Violet} & $0.068\pm0.0082 $ & 0.044 & $ 0.12\pm0.034 $ &  0.13 & $0.093\pm0.011 $ & 0.059 & $  0.30\pm0.083 $ &  0.31 & $  3.2\pm 0.97 $ & $  5.2\pm  1.7 $  \\
\textit{Blue} & $  1.5\pm0.043 $ &   1.5 & $  1.7\pm 0.11 $ &   1.7 & $  2.1\pm0.058 $ &   2.1 & $  4.1\pm 0.26 $ &   4.1 & $  1.9\pm 0.13 $ & $    2\pm 0.14 $  \\
\textit{Green} & $ 0.95\pm0.032 $ &   1.0 & $  2.1\pm  0.1 $ &   2.2 & $  1.3\pm0.043 $ &   1.4 & $  5.2\pm 0.24 $ &   5.4 & $   4.0\pm 0.23 $ & $   4.0\pm 0.22 $  \\
\textit{Yellow} & $ 0.13\pm0.0096 $ &  0.11 & $ 0.17\pm0.035 $ &  0.15 & $ 0.18\pm0.013 $ &  0.15 & $ 0.41\pm0.084 $ &  0.36 & $  2.3\pm 0.49 $ & $  2.4\pm 0.59 $  \\
\textit{Red} & $ 0.66\pm0.024 $ &  0.66 & $    1.0\pm0.077 $ &  0.95 & $  0.90\pm0.033 $ &   0.90 & $  2.4\pm 0.19 $ &   2.3 & $  2.7\pm 0.23 $ & $  2.6\pm 0.23 $  \\\hline
    \end{tabular}
\tablefoot{Line luminosities are corrected for lensing magnification.}
\end{table*}

\subsection{Disentangling the dust continuum}
\label{sect:continuum_model}

\begin{figure*}
\begin{center}
\includegraphics[width=\textwidth]{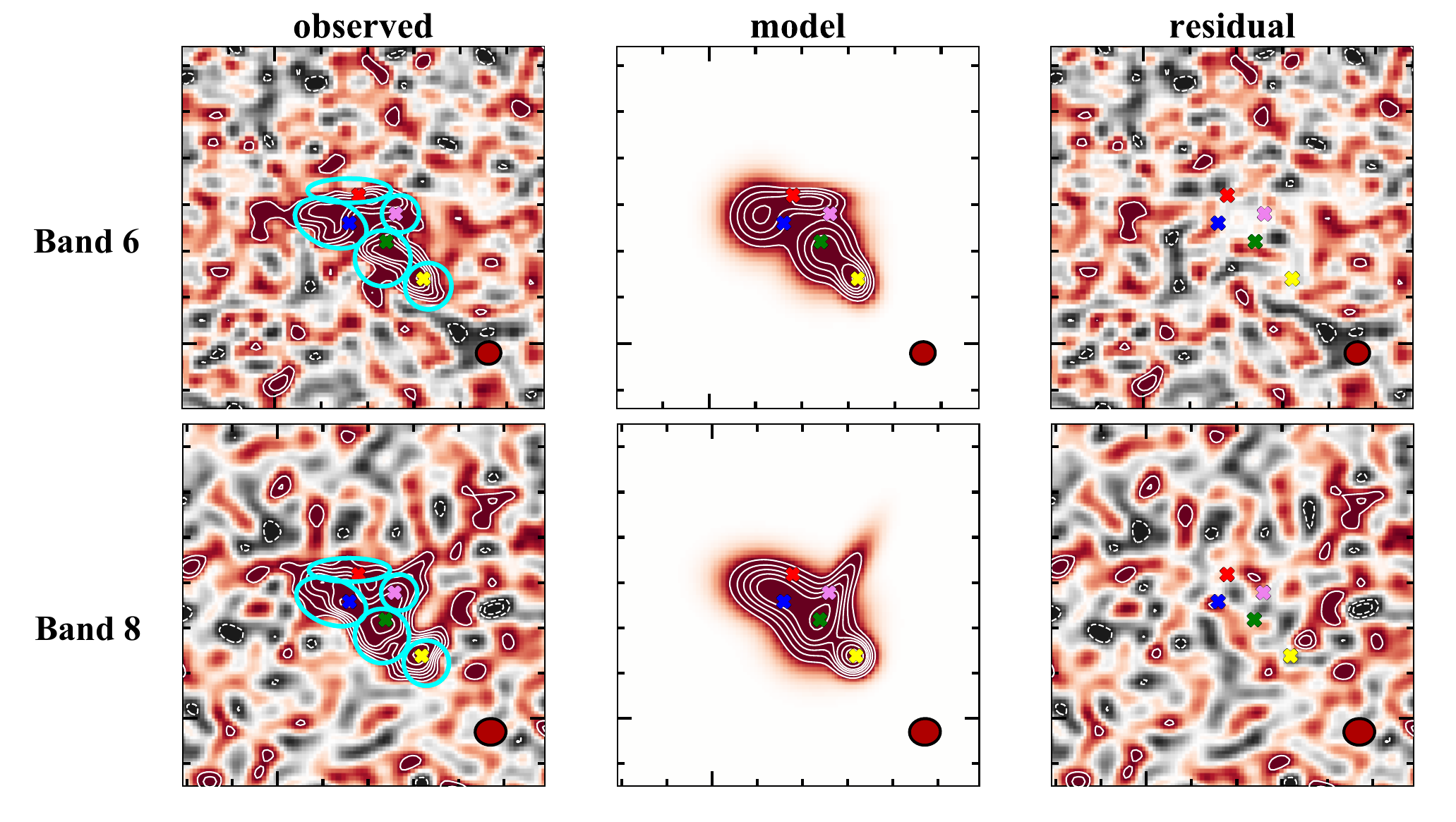}
\end{center}
\caption[]{
Decomposition of the band 6 (about $158\,\mu$m, \emph{top}) and band 8 (about $88\,\mu$m, \emph{bottom}) dust continuum for the five components. 
The $4''\times4''$ cutouts of the observed, {\sc imfit} model, and residual maps are presented from left to right. 
The solid and dashed contours show the intensity at the 2, 3, 4, 5, 6, 8, and 10$\sigma$ levels and $-3$ and$-2\sigma$ levels, respectively, which are measured based on the residual map. 
The spatial position of the five \oiii\ components are marked with the color crosses. 
The cyan circles indicate the elliptical apertures that are used for the aperture photometry. 
\label{fig:cont_model}}
\end{figure*}

We also estimated the flux densities of the continuum emission in band 6 and band 8 for each component. 
We first performed a forced-aperture photometry. 
Given that \oee shows a better correlation than \cofe \ with the SFR  \citep[e.g.,][]{delooze14,olsen17}, we optimized spatial positions and shapes of the apertures following the five components of the \oiii model. 
In the left panel of Figure~\ref{fig:cont_model}, we show the five aperture positions and shapes and summarize the extracted flux densities in Table~\ref{tab:dustcomponents}. 
In this approach, the apertures do not overlap with each other; however, we cannot rule out the possibility that some measurements within the aperture are affected by contributions from other components. 

Next, we modeled and decomposed each contribution by using CASA {\sc imfit}. 
From the observed morphology, we assumed four components. The violet (red) component was not included in the fitting for the band 6 (band 8) maps, but we extracted its photometry with a radius of $0\farcs2$ aperture on the residual map. 
We fixed the spatial positions of the faint components of red and violet when they are included in the model. We kept the flux density as a free parameter, while we fixed some of the other parameters (size, axis ratio, and PA) in a routine process to obtain stable results. 
In Table~\ref{tab:dustcomponents}, we summarize the flux density estimates in the both bands for each component. 
By assuming a single modified blackbody and a spectral index of $\beta=1.7$ in a spatially averaged manner \citep{akins22}, we also derived the dust temperature ($T_{\rm dust}$) and  IR luminosity ($L_{\rm IR}$). 
Due to the low S/N for the violet and red components, we derived their $L_{\rm IR}$ values by fixing their $T_{\rm dust}=41$~K taken from the spatially integrated best-fit value presented in \cite{akins22}. 
These estimates are also listed in Table~\ref{tab:dustcomponents}. 
The sum of the five components gives $0.45\pm0.20$\,mJy and $1.71\pm0.58$\,mJy in bands 6 and 8, respectively. 
The band 8 continuum is comparable to the total flux estimate in Table~\ref{tab:OIIIcomponents}, while the band 6 continuum model underestimates the total flux by $\sim50\%$. 
This indicates that the diffuse continuum emission, significant in band 6 \citep{akins22}, is not fully recovered by this decomposition analysis.
The continuum model and residuals are shown in Fig.~\ref{fig:cont_model}. We note that we here use a higher resolution weighting scheme compared to \citet{akins22} and thus the difference is less evident in the residuals.

\section{Discussion}
\label{sect:disc}

\subsection{Resolved and overall physical properties}
\label{sec:decomp2}

Based on the component analysis from the previous subsections, we estimate the line luminosity and the continuum properties of each component.  
The line luminosity is estimated both using the observations and the model results. In Table~\ref{tab:resolved_linelum} we provide the line luminosity derived for each component.  The uncertainty is estimated from the observed data, but does not include the uncertainty on the absolute flux calibration.  

Based on our resolved line and continuum estimates, we can evaluate physical properties of the full source and the individual components. 
While blue and red are further decomposed into two subcomponents in Sect.~\ref{subsect:linecontdecomp}, here we focus on the physical properties of the brighter component that dominates their total emission. 

\subsubsection{Total SFR}
We calculated the total SFR by summing the obscured (SFR$_{\rm IR}$) and unobscured SFR (SFR$_{\rm UV}$) based on the ALMA and HST observations. 
We measured the UV luminosity $L_{\rm UV}$ for blue and green components with a $0\farcs2$-radius circular aperture placed at the peak positions of the two blobs observed in the HST maps. 
Because of the lack of the HST counterparts, we obtained a 3$\sigma$ upper limit with the same size aperture for the other components. 
We used the same conversions from IR and UV luminosity to SFR as \cite{akins22} 
based on the calibration presented in \citet{murphy11}
and obtain a total SFR of $16.5_{-6.8}^{+22}\,M_{\odot}$\,yr$^{-1}$. 
We did not account for the calibration uncertainty in the error estimate.

We confirm  consistency with previous measurements \citep{bakx21,akins22} given the recovery fraction of the total dust continuum in our measurement (Sect.~\ref{sec:decomp2}). 
We found that the obscured fraction $f_{\rm obs}$ is $\sim80$\% in the blue and green components, respectively, and more than $\sim90$\% in the other components, indicating the high obscuration in every part of the system. 
We summarize SFR$_{\rm IR}$, SFR$_{\rm UV}$, and $f_{\rm obs}$ for the individual components in Table~\ref{tab:comp_dynamical_mass}. 

\subsubsection{SFR--$L_{\rm line}$ relation}
\begin{figure*}
\includegraphics[width=\columnwidth]{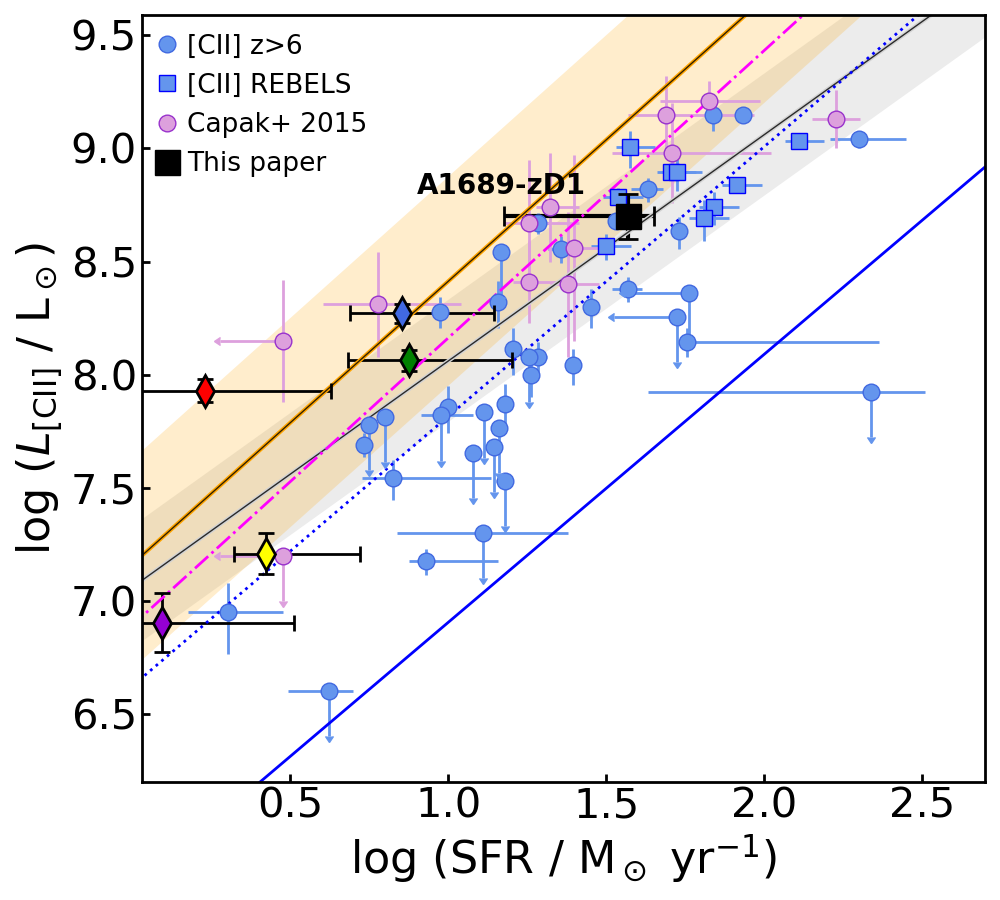}
\includegraphics[width=\columnwidth]{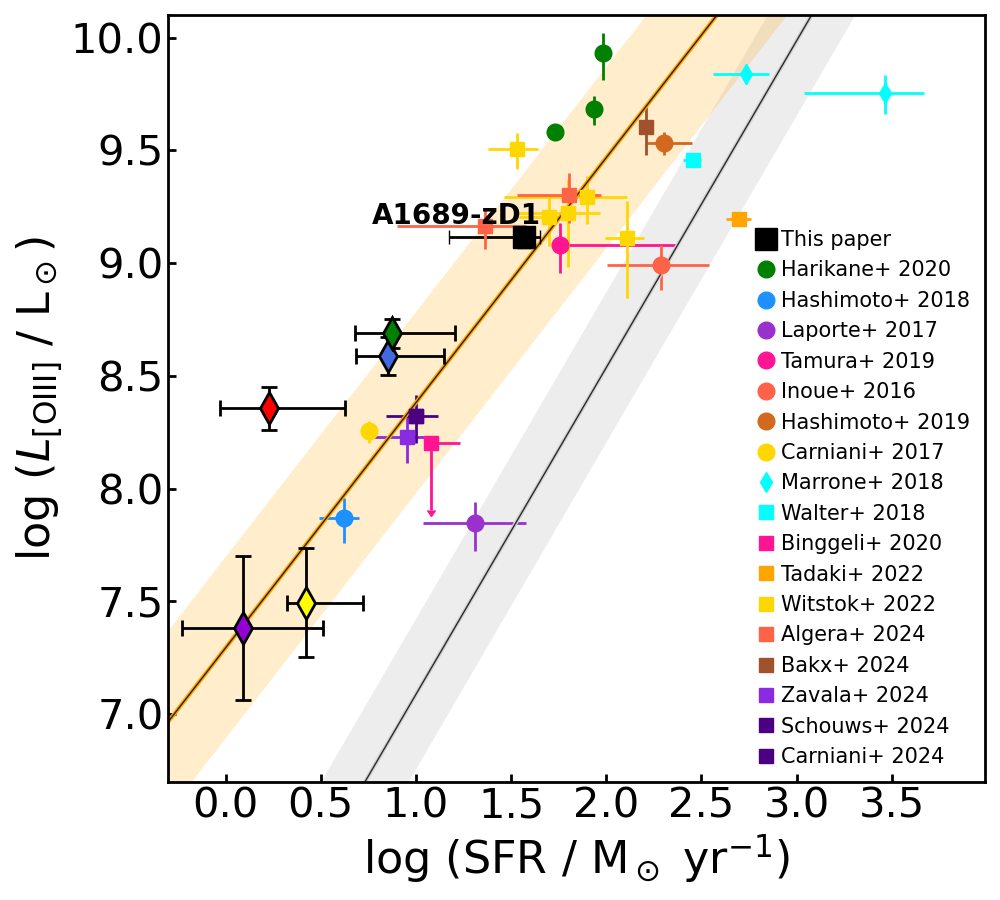}
\caption[]{
Line luminosity vs. SFR for $L_{\rm [CII]}$ (\textit{left}) and $L_{\rm [OIII]}$ (\textit{right}). Black outline diamonds with color centers indicate the color-coded individual components from Sect.~\ref{subsect:linecontdecomp}. The uncertainty on the SFR-axis for each component corresponds to the range given by a $T_{\rm dust}=25-55$\,K. The orange and gray lines and associated regions show the relations for star-forming galaxies and low-metallicity dwarf galaxies from \citet{delooze14}.  Data from the literature is included for comparison for high-$z$ galaxies without AGNs \citep{capak15,inoue16,knudsen16,pentericci16,bradac17,laporte17,carniani17,smit18,walter18,hashimoto18,hashimoto19,tamura19,laporte19,harikane20,binggeli20,fujimoto21,laporte21,molyneux22}. We note that the data from the REBELS survey include the sources with $L_{\rm [CII]}$ and UV+IR SED-estimated SFR \citep{sommovigo22,ferrara22}. On the right panel, for \oiii we also include the results from \citet{marrone18}.
In the left panel, the blue and magenta lines represent model predictions from \citet{vallini15} and \citet{olsen17}.}
\label{fig:linesfr}
\end{figure*}

In Fig.~\ref{fig:linesfr} 
we show SFR--$L_{\rm [CII]}$ and SFR--$L_{\rm [OIII]}$ relations for the individual components and the entire system of A1689-zD1. We also show for comparison results obtained in local galaxies and $z>6$ galaxies without active galactic nuclei (AGNs) from the literature. 
For A1689-zD1, we find that the relation in both lines is consistent with previous results for $z>6$ galaxies, where the \cii line is slightly below the relation from local galaxies, but within its 1$\sigma$ range; whereas the \oiii line agrees well with it. 
For individual components, we find that the relation has a little scatter, but still falls within the 1$\sigma$ range of the local galaxies. These results are consistent also with ALMA results for $z>6$ lensed galaxies \citep[e.g.,][]{knudsen16,fujimoto21,molyneux22}. 

Given that the emission is resolved, it is also natural to compare this to local galaxies. For example, the SHINING survey of nearby star-forming galaxies extensively studied FIR fine-structure lines and compared this to continuum properties \citep[e.g.,][]{herreracamus18}. Comparing line luminosity to the FIR luminosity and the FIR surface brightness, $\Sigma_{\rm FIR}$, of A1689-zD1 to the results from the SHINING, we note that \cii results are within the scatter and consistent with that of SHINING results, while the \oiii line appear about a factor two brighter than the brightest \oiii detections in the SHINING survey.

\begin{figure}
    \centering
    \includegraphics[width=0.99\columnwidth]{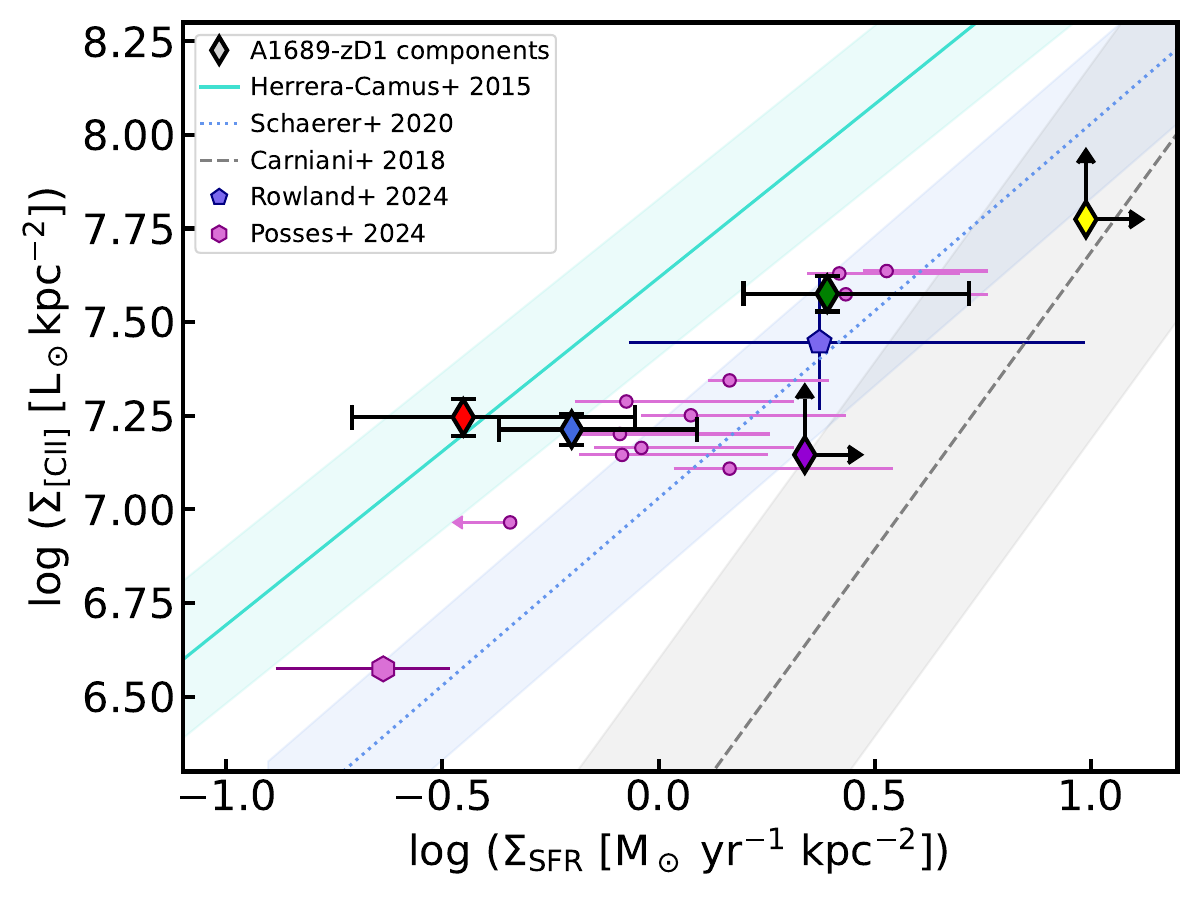}
    \caption{
    Surface brightness, SFR, \cii, $\Sigma_{\rm SFR}$, and $\Sigma_{\rm [CII]}$ for the individual components of  A1689-zD1 compared to recent results of the resolved high-$z$ galaxies REBELS-25 \citep{rowland24} and CRISTAL-05 \citep[][the circles show resolved measurements across the galaxy]{posses25}.  The turquoise line and region shows the local relation from \citet{herreracamus15}, the blue dotted line and region show the relation from ALPINE  \citep[][assuming an average area of 8\,kpc$^{2}$]{schaerer20}. The dashed gray line and region show the relation for $z=5.5-7$ galaxies \citep{carniani18}.}
    \label{fig:sigSFR_sigCII} 
\end{figure}

In a compilation of $z=5-7$ galaxies with \cii\ detections, \citealt{carniani18} find that "normal" star-forming galaxies deviate from the local empirical relation of $\Sigma_{\rm SFR}-\Sigma_{\rm [CII]}$ with a typically lower \cii\ surface brightness. Comparing the SFR-$L_{\rm [CII]}$ relation from the ALPINE survey \citep{schaerer20}, assuming average extended emission of 8\,kpc$^{2}$ \citep[as done in ][]{posses25}, it is positioned between the relation from \citet{carniani18} and the local relation from \citet{herreracamus15}. Differences between the different populations (e.g.,\ local vs. high-redshift galaxies) could be caused by heating mechanisms, effects of metallicity, or the ionization parameter \citep[e.g.,][]{carniani18}.  In Fig.~\ref{fig:sigSFR_sigCII}, we compare the resolved \cii emission of A1689-zD1 with those relations and with recent results of the two resolved high-$z$ galaxies REBELS-25 \citep{rowland24} and CRISTAL-05 \citep{posses25}.  For REBELS-25, we estimated $\Sigma_{\rm SFR}$ and $\Sigma_{\rm [CII]}$ assuming a 2D Gaussian distribution and the effective radius from \citet{rowland24} for estimating the area. For the A1689-zD1 components, we also assumed a 2D Gaussian distribution with the model FWHM from Table~\ref{tab:CIIcomponents}. We  note that the plotted errors do not include the uncertainties of the sizes.  Given the large uncertainties on the SFR estimates, the individual components of A1689-zD1 are comparable to the high-redshift results of REBELS-25, CRISTAL-05, and the relation from ALPINE. The red component appears partly consistent with the local relation from \citet{herreracamus15}, which could be indicative of a more efficient heating \citep[e.g.,][]{herreracamus15}. 

\subsubsection{SFR--$L_{\rm [OIII]}/L_{\rm [CII]}$ relation}

\begin{figure}
    \centering
    \includegraphics[width=0.99\columnwidth]{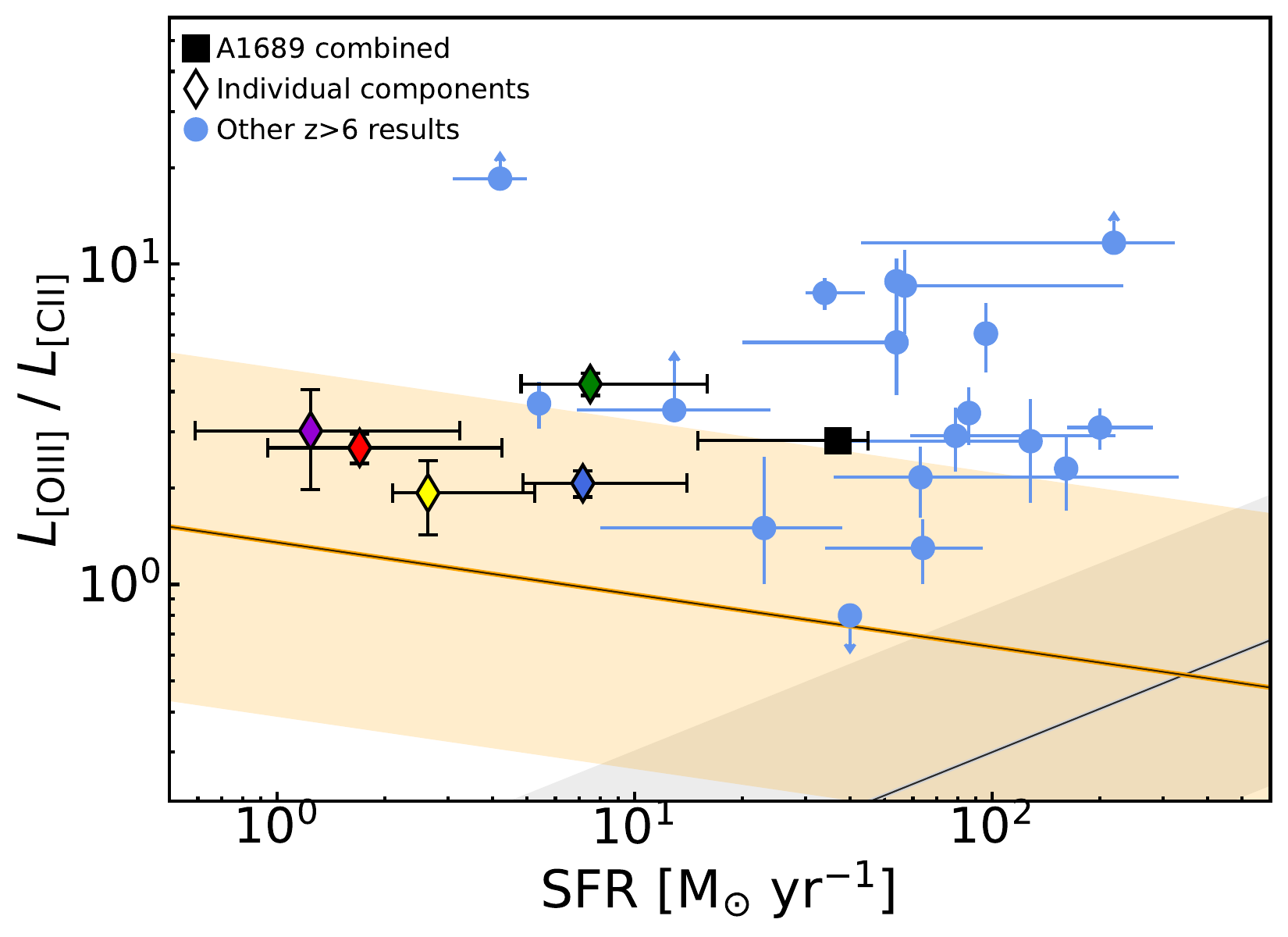}
    \caption{Line luminosity ratio of \oiii and \cii vs. SFR for each component of A1689-zD1 compared to other high-$z$ detections \citep{inoue16,laporte17,carniani17,hashimoto18,hashimoto19,tamura19,laporte19,harikane20,witstok22,algera24,bakx24}.  The diamonds represent the individual components color-coded according to their name, and the star formation is the one derived from the aperture measurements, as given in Tables~\ref{tab:resolved_linelum} and \ref{tab:dustcomponents}. The orange and gray region and line indicate the range of the line ratio for local low-metallicity dwarf galaxies and starburst galaxies from \citet{delooze14}. 
    \label{fig:lineratio} } 
\end{figure}

We also compared the line luminosity ratio of \cii and \oiii to the SFR for $z>6$ galaxies, shown in Fig.~\ref{fig:lineratio}.  
While the estimates of the line ratios of the components span a range of a factor two, they are relatively consistent with the overall result for A1689-zD1. It has been found that $L_{\rm [OIII]}/L_{\rm [CII]}$ is typically above 2 for $z>6$ galaxies, however, this is based on a relatively small number of sources so far \citep{harikane20,Carniani2020}. High $L_{\rm [OIII]}/L_{\rm [CII]}$ ratios have also been found in local low-metallicity dwarf galaxies \citep[e.g.,][]{cormier15}. However for local luminous IR galaxies, the ratio tends to be $\leq 1$ \citep[e.g.,][]{howell10,diazsantos17}. 

Using a compilation of new data and literature results, \citet{harikane20} modeled the line ratios. They use simple photoionization models and discuss how several different physical effects impact the ratio; for example, the metallicity, ionization parameter, gas density, C/O ratio, and cosmic microwave background (CMB) attenuation.  They find different properties can explain this, in particular very high ionization parameters or low photo-dissociation region coverage. Extensive modeling using cosmological radiation-hydrodynamics simulation have been published by \citet{katz22}.  Those simulations show that several effects, such as ionization parameter, CMB attenuation, and the ISM gas density, can play a significant role; however, in order to reproduce the observed results, the C/O abundance need to be lower than the solar value \citep[e.g.,][]{bakx24}. 

Comparing to the line ratios of the SHINING survey for local star-forming galaxies \citep{herreracamus18}, we note that the \oiii\!-to-\cii luminosity ratio is about a factor two higher. Given the relatively well-sampled global spectral energy distribution \citep[SED;][]{bakx21}, we estimate the continuum ratio $S_{\rm 63\mu m}/S_{\rm 122\mu m} \sim 1.2-1.7$. In the SHINING sample, the sources with $S_{\rm 63\mu m}/S_{\rm 122\mu m} > 1$ are typically also the sources with $L_{\rm [OIII]}/L_{\rm [CII]} > 1$; however, we  note that SHINING  only has few sources with such high line ratios.

\subsubsection{Size-velocity relation}

\begin{figure}
    \includegraphics[width=\columnwidth]{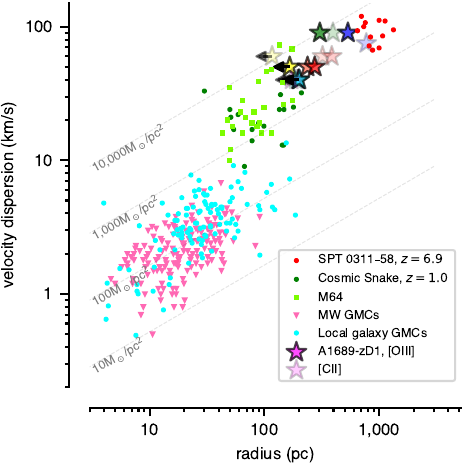}
    \caption{Radius-velocity relation of the clumps in A1689-zD1 compared to giant molecular clouds (GMCs) in the Milky Way  \citep{Heyer2009}, in the nearby spiral M64 \cite{Rosolowsky2005}, a compilation of other low redshift galaxy GMCs \citep{Bolatto2008}, in the  so-called Cosmic Snake at $z=1.0$ \citep{dessauges19}, and in the clumpy, lensed  SPT\,0311$-$58 at $z=6.9$ \citep{spilker22}. Lines of constant surface mass density are shown, following \citet{spilker22}. The clumps in A1689-zD1 have a high surface mass density, 1.5\,dex higher than low-$z$ GMCs and similar to SPT0311$-$58 and the Cosmic Snake, though with clump sizes that are factors of several smaller than in the more massive galaxy SPT\,0311$-$58 and much larger than observed in less star-forming galaxies at lower redshift. Our smallest components are unresolved, indicating the possible presence of even smaller clumps, possibly corresponding in size to those in the Cosmic Snake or M64.
    \label{fig:vel_radius} } 
\end{figure}

In Fig.~\ref{fig:vel_radius} we also compare the relation between size and velocity dispersion of the individual components with other spatially-resolved measurements from local to high-redshift galaxies. 
For size, we adopted the effective radius for the comparison, assuming for the Gaussian profiles that the half-width-at-half-maximum is comparable with the effective radius. 
By assuming virialized clouds in the same manner as \citet{spilker22}, we also calculated the expected size and velocity dispersion at given mass surface density (illustrated by the dotted lines in Fig.~\ref{fig:vel_radius}).  

We find that the individual components are different from the distribution of the clumps observed in the local galaxies, but fall between the parameter spaces observed in the high-redshift, lensed, star-forming galaxies of the ``Cosmic Snake'' at $z=1.036$ \citep{dessauges19} and SPT\,0311$-$58 at $z=6.900$ \citep{spilker22}. 
The distribution of the individual components are aligned with the dotted line, showing that their mass surface densities are comparable with those of the star-forming clumps observed in the other high-redshift lensed galaxies. 
This might indicate that the internal star-forming mechanisms resemble each other, even though there is a large difference in the total SFRs of these high-redshift lensed galaxies, from a few tens to thousands of solar masses per year. 

\subsubsection{Kinematics}
\label{sec:kin}

\begin{figure*}
\includegraphics[width=0.50\textwidth]{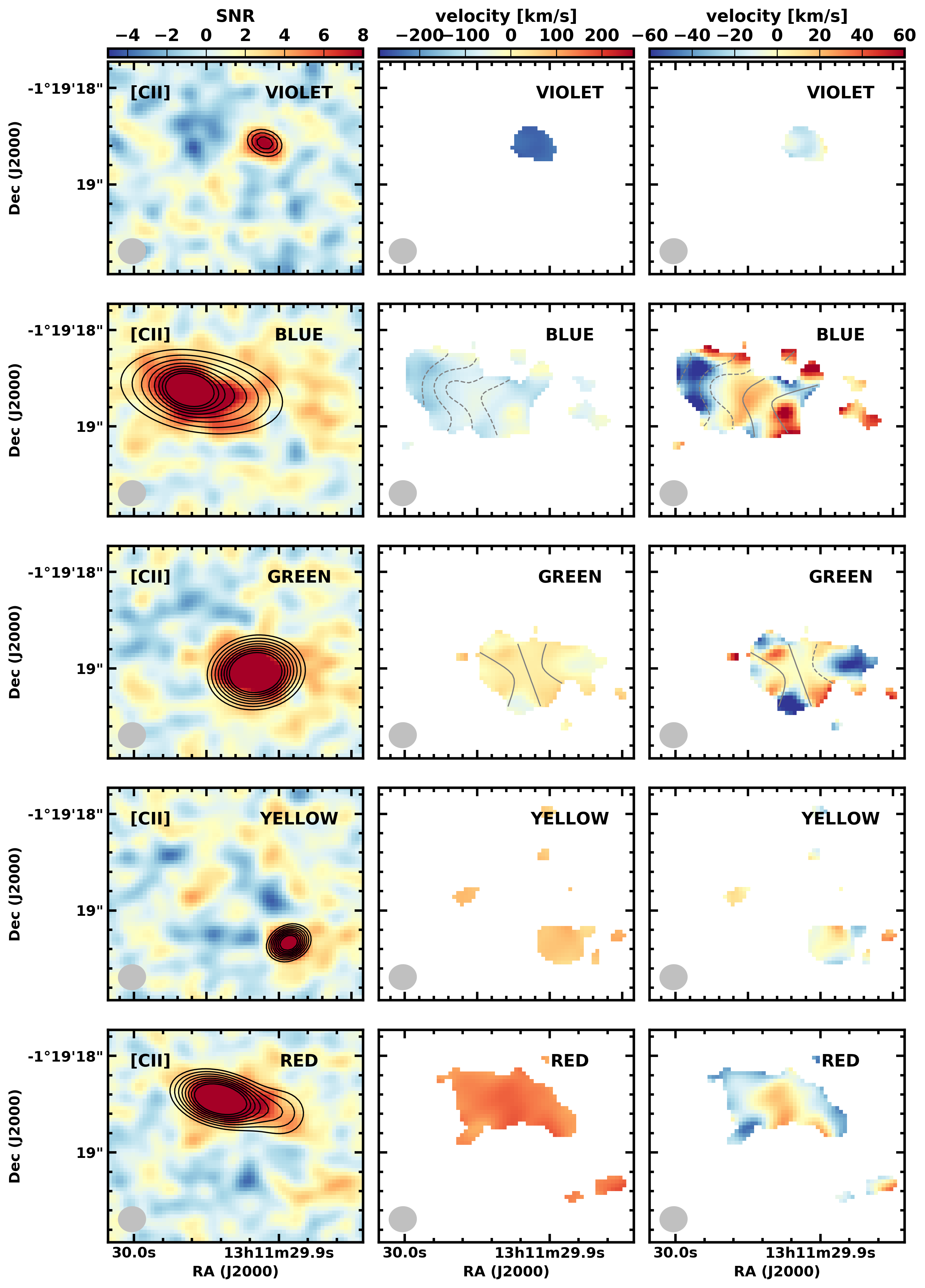}
\includegraphics[width=0.50\textwidth]{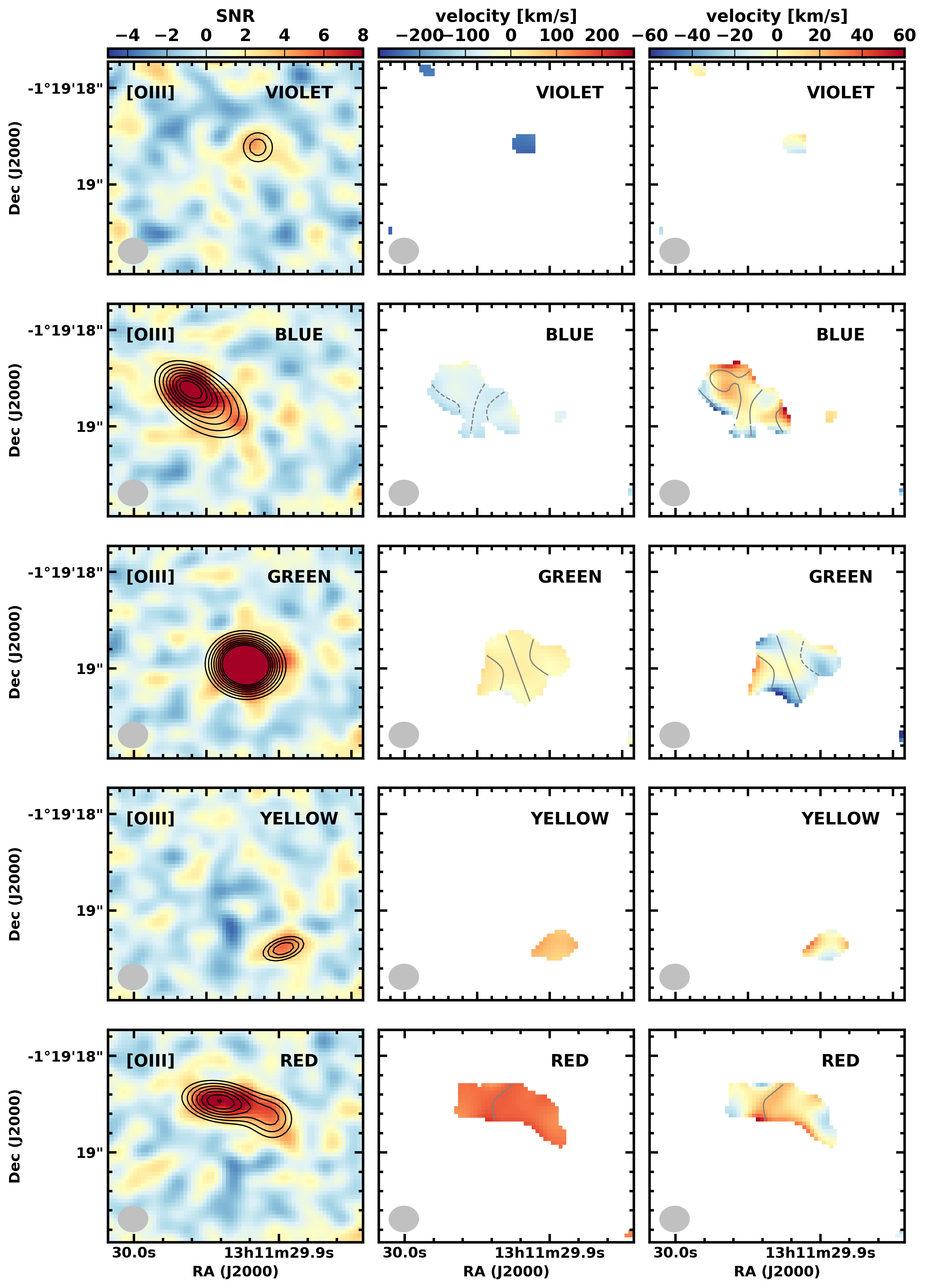}
\caption[]{Moment-0 and moment-1 maps for each component: \cii results (left) and \oiii results (right). The models of the other components have been subtracted from the data cube. The left columns show the moment-0 maps integrated over the line of the component; the middle columns the moment-1 given relative to the A1689-zD1 overall velocity field with zero velocity corresponding to $z=7.133$; and the right columns the moment-1 map for each individual velocity (same velocity used for both \cii and \oiii).  The overlaid contours represent the fit models, in the moment-0 maps these correspond to 3, 4, 5, 6, 7, 8, 9, and 10$\sigma$; in the moment-1 maps, these correspond to velocity in km\,s$^{-1}$ in steps of 10\,km\,s$^{-1}$. The moment-0 maps are given in S/N, where the rms in each map is measured excluding the source region (for \cii: rms = 5.4, 11.6, 11.5, 6.4, and 8.3 mJy\,km\,s$^{-1}$\,beam$^{-1}$, respectively, and for \oiii: rms =  27.7, 50.8, 54.1, 29.4, and 38.1  mJy\,km\,s$^{-1}$\,beam$^{-1}$, respectively (from top to bottom).   
\label{fig:components_imm01}}
\end{figure*}

To analyze the kinematics of each component, we created a fiducial 3D data cube for each component by subtracting the 3D models of the other components (Sect.~\ref{subsect:linecontdecomp}) and produced moment-0 and moment-1 maps. 
In Fig.~\ref{fig:components_imm01}, we show the moment-0 and moment-1 maps of both the \cii and \oiii lines for each component.
The moment-0 maps are shown in terms of their S/N, with the rms indicated in the figure caption. 
Two versions of the moment-1 are displayed: one showing the velocity relative to the overall galaxy ($z=7.133$) and one with zero velocity centered in the component.   

We find in the \cii moment-1 map that blue component shows a symmetric velocity structure, from blueshifted in NE to redshifted in SW around its intensity peak. Apart from the blue component, we do not find clear velocity gradients in the moment-1 maps of any other component in the \cii or \oiii. 
We thus regarded blue as a potential rotator and the other components as dispersion-dominated systems. 
We note, however, that blue does not show a similar velocity gradient in the \oiii\ moment-1 map. 
As discussed in \cite{rizzo21}, 
this may be explained by the difference in the motions between ionized and cold neutral gases, where the ionized gas is more likely tracing turbulent motions affected by stellar feedback for example, although the rotation has been seen in \oiii\ at $z \sim 9$ \citep{tokuoka22}. In fact, \citet{akins22} report different line profiles between \cii\ and \oiii\ at the highest SFR surface density region, mostly corresponding to green, which may be explained by the outflow of the ionized gas. 

\subsubsection{Dynamical masses}

We also estimated the dynamical masses, $M_{\rm dyn}$, of the individual components. 
For blue, we assumed a disk-like gas distribution, given the potential rotation (Sect.~\ref{sec:kin}), and followed Equation~4 in \citet{dessauges20}: 
\begin{eqnarray}
\left (\frac{M_{\rm dyn}}{M_{\odot}} \right) &=& 1.16\times10^{5} \left(\frac{v_{\rm circ}}{\rm km\,s^{-1}}\right)^{2} \left(\frac{2r_{\rm e}}{\rm kpc} \right),  
\end{eqnarray}
where $r_{\rm e}$ is the effective radius, $v_{\rm circ}$ is the circular velocity of the gaseous disk, including the inclination correction, given by $v_{\rm circ}=1.763\sigma_{0}/{\rm sin}(i)$ and $i={\rm cos^{-1}({\rm FWHM_{2}/FWHM_{1}})}$. 

For the other components, we assume 
dispersion-dominated systems and virialized bodies, following Equation~10 of \citet{bothwell13}, as 
\begin{eqnarray}
\left (\frac{M_{\rm dyn}}{M_{\odot}} \right) &=& 1.56\times10^{6} \left(\frac{\sigma_{\rm 0}}{\rm km\,s^{-1}}\right)^{2} \left(\frac{r_{\rm e}}{\rm kpc} \right) , 
\end{eqnarray}
 \noindent 
 where $\sigma$ is the line-of-sight velocity dispersion and $r_e$ is the effect radius of the sources as described from the fitting above, and thus giving the dynamical mass within the effective radius.
Because the \cii and \oiii lines could trace different gas motions and regions, we estimated two $M_{\rm dyn}$ values for each component separately --- \cii ($M_{\rm dyn, [CII]}$) and \oiii\ ($M_{\rm dyn, [OIII}$). We note that given the complicated morphology and kinematics of the system, each component might not be virialized. Nonetheless, the estimates provide a meaningful estimate of the dynamical mass. 
We summarize the various estimates of $M_{\rm dyn}$ in Table~\ref{tab:comp_dynamical_mass}.

\begin{table}
    \centering
    \caption{Intrinsic physical properties for the different components.}
    \label{tab:comp_dynamical_mass}
    \setlength{\tabcolsep}{4pt}
    \begin{tabular}{@{}lccccc@{}}
\hline
\hline
        Comp- & SFR$_{\rm IR}$    & SFR$_{\rm UV}$  & $f_{\rm obs}$ & ${
        \rm M}_{\rm dyn, [OIII]}$ & ${\rm M}_{\rm dyn, [CII]}$   \\
         onent    &  [M$_\odot$~yr$^{-1}$] &  [M$_\odot$~yr$^{-1}$] & [\%] & [10$^{9}$M$_\odot$] & [10$^{9}$M$_\odot$] \\[3pt]
\hline\\[-9pt]
    \textit{Violet} & 1.2$_{-0.6}^{+2.0}$ & $<0.2$ & $> 86$ & 0.5 &\nodata \\[3pt]
    \textit{Blue}   & 5.4$_{-2.3}^{+6.7}$ & 1.7 $\pm$ 0.1 & 76 & 5.9 & 5.7      \\[3pt]
    \textit{Green}  & 5.7$_{-2.7}^{+8.4}$ & 1.7 $\pm$ 0.1 & 77 & 4.3 & 6.0  \\[3pt]
    \textit{Yellow} & 2.6$_{-0.5}^{+2.5}$ & $<0.2$ &$> 93$ & \nodata & \nodata \\[3pt]
    \textit{Red}    & 1.6$_{-0.7}^{+2.4}$ & $<0.2$ &$> 89$ &  1.8 & 2.9 \\[3pt] \hline\\[-9pt]
    Sum   & 16.5$_{-3.7}^{+11.4}$ & 3.4 $\pm$ 0.1 & 83 & 12.5 & 14.6 \\[3pt] 
\hline
    \end{tabular}
\end{table}

\subsection{Mass of A1689-zD1}

While A1689-zD1 was originally discussed as a modest, star-forming galaxy \citep{bradley08,watson15}, there are now several lines of evidence that indicate the galaxy is much larger and more massive than the stellar and gas mass derived from the continuum emission. 
The 12\,kpc extent of the \cii emission \citep[regardless of its origin;][]{akins22} as well as the increased IR SFR \citep[][and Sect.~\ref{sect:obs}]{bakx21,akins22} point to a bigger system. Furthermore, the dynamical mass estimates of the clumps (Sect.~\ref{tab:comp_dynamical_mass}) indicate that the galaxy may be much more massive than originally believed, with a summed dynamical mass about $1.5\times10^{10}\,M_\odot$. This is an order of magnitude larger than the estimated stellar mass based on SED modeling including \emph{Spitzer}-IRAC $3.6\,\mu$m and $4.5\,\mu$m detections. This implies an unfeasible high gas mass fraction given the approximately solar metallicity inferred from the FIR emission line ratios \citep{killi23}. The total metal mass of the galaxy would then exceed what could be produced the core collapse supernovae in a Chabrier initial mass function. On this basis, the gas mass fraction of this large dynamical mass is unlikely to be more than about 50\%. 
There is certainly sufficient dust in the galaxy: a screen of half the estimated dust mass in A1689-zD1 distributed uniformly across 2\,kpc$^2$ would result in an $A_V\sim40$\,mag using the Milky Way dust extinction to column density ratio \citep{watson2011}. Since the star-formation obscuration fraction is about 90\% \citep{akins22}, the stellar mass is unlikely to be underestimated by more than this, namely,\ by more than an order of magnitude. The stellar mass is therefore likely to be about 1--2$\times10^{10}\,M_\odot$. These issues are discussed in greater depth in \citet{killi23}. 

\subsection{Considering the physical origin of A1689-zD1}

The complex gas structure and kinematics of A1689-zD1 revealed by the ALMA
observations are different and exhibit a higher degree of irregularity compared to
other $z>7$ galaxies, where the multiplicity is often characterized by
two or three \cii components \citep[e.g.,][]
{matthee17,carniani17,hashimoto19}, with separations on
scales of 2--7\,kpc. Other observations show either \cii that is separated from other components (e.g.,\ \oiii) and extended \citep[e.g.,][]{maiolino15,marrone18}. Single-components at scales of about
3\,kpc with some degree of rotation have also been claimed in UV-bright galaxies at
$z>6.5$ \citep[e.g.,][]{smit18,tokuoka22}. Other \cii detections either do not have
sufficient S/N or angular resolution to resolve the
emission \citep[e.g.,][]{knudsen16,bradac17,pentericci16}.

Many recent results for galaxies at $z>4$, particularly from ALMA observations, show very clear velocity gradients \citep[e.g.,][]{jones17,smit18,neeleman20,tokuoka22,rowland24,posses25}. In many cases, the observed velocity gradients are interpreted as gas rotation and, in some cases, this can also be aptly described by rotational models 
\citep[e.g.,][]{carniani13,litke19,jones21,rizzo21}. 
These results are based both on non-lensed sources and gravitationally lensed galaxies and also across a range of different luminosities and masses. Certainly, the same system can change appearance due simply to inclination angle or evolution over a relatively short time. Only a few hundred million years separating $z=7$ from $z=5-6$ would be sufficient for a system to change from disk rotation to merger-dominated, according to the simulations of \citet{kohandel19}. 

It is worth noting that our angular resolution is about 5-20 times better than galaxies studied by \citet{smit18} and \citet{matthee17} due to the observational setup and the gravitational lensing magnification, which  allows us to clearly rule out a simple ordered rotation for A1689-zD1. While a somewhat lower resolution would likely be sufficient to demonstrate that A1689-zD1 is not fully explained based on simple rotation, the sensitivity of data we have here, combined with a factor $\sim$9.6 in gravitational magnification reveal structures that are rarely detected in galaxies at high redshifts. Our conclusion here is somewhat at odds with \citet{Wong2022}, who argued for a clear velocity SE-NW across the galaxy, with the NW being more dust-obscured. However, the presence of the violet component co-spatial with the redder part of the galaxy complicates a simple merger picture, as does the possible rotation and extent of the blue component. The results for the lensed galaxy SPT\,0311$-$58 at $z=6.9$ observed with ALMA, has a comparable S/N and sub-kiloparsec resolution and also displays a clumpy structure \citep{spilker22}. This suggests that a high S/N combined with spatial resolution significantly below the kpc scale may be important in fully exploring the dynamical state of at least some $z>6$ galaxies. 

\subsubsection{An early galaxy assembly possibly consisting of five small clumps}

One possible interpretation of the detection of several components within projected distances of $\sim3$\,kpc and the detection of extended \cii emission is that we are witnessing the early assembly of the inner kiloparsec-scale region of a galaxy. In such a scenario each of the components would represent ISM/mass substructure, possibly as (super)giant molecular clouds or young stellar clusters in the assembly of the bulge region. Given the large relative offsets and velocities, the ordered motion seen in disk galaxies would not have been established thus far.  
The extended \cii detection \citep{akins22}, would either be halo gas or possibly part of extended ISM, part of a bound system. If this was the case, we might expect an extended stellar component following the structure of the extended \cii, which could be detectable with JWST. As argued above, the stellar mass may be much larger than indicated by the \emph{Spitzer} observations. However, simulations of early galaxies suggest that they are unlikely to have effective radii larger than 2--3\,kpc \citep{roper22}, indicating that such an extended galaxy interpretation is less favored. This suggests that if A1689-zD1 has a large obscured stellar mass, it is most likely concentrated in the inner few kiloparsecs.

\subsubsection{Galaxy merger interpretation}

The blue, green, and red components dominate the galaxy's emission. As discussed above (and essentially indicated by \citealt{Wong2022}), it is possible that the blue and green components could be associated with the two rest-frame UV components detected with HST. Given the velocity difference of 80\,km\,s$^{-1}$ and projected physical separation of 0.8--1.4\,kpc between these two (the range is given by the slightly different positions in the \cii and \oiii data), one interpretation could be interaction and merger.  The red component shows an extended tail-like structure, which could suggest that this is an additional component being pulled towards the green and blue, while the elongated structure represents a tidal tail of the gas. The red component is separated by 0.7--1\,kpc in projection and 220\,km\,s$^{-1}$ from the blue component and by 1.8-2.1\,kpc in projection and 140\,km\,s$^{-1}$ from the green. This could suggest a scenario of a merger event between green and blue, with red being a gas-rich structure or a scenario of a triple-merger, where the red component contains a more dust-obscured stellar population and a tidal disruption of the gas distribution.  
The sizes and distribution could be consistent with the `merger' phase observed in the simulations of \citet{kohandel19}.  
However, without the spatial information from the radial direction, such merger scenarios are difficult to constrain.

\subsubsection{Combined scenario}

In the two subsections above, we describe the assembly of the inner few kiloparsecs of a galaxy or a merger scenario of two or three galaxies. A third plausible interpretation could be a combination of the two. Given the complex, almost chaotic structure, blue, green, and yellow appear to align in one direction and could represent the inner region of a galaxy, while red and violet could comprise more galaxy components that are interacting and possibly being accreted onto the main galaxy structure. The much higher dust obscuration of these NW components may support such a scenario.

Even though the ALMA data is of exceptional quality and reveals substructure so far unmatched in high-$z$ galaxies, two pieces of information are missing: the radial spatial distribution, which cannot be measured, and the distribution of the stellar mass. This latter may well be obtained through deep observations with JWST. Zoom-in numerical simulations are necessary to provide additional constraints, similar to those of \citet{kohandel19}.

\subsection{Different star-formation indicators in the galaxy}

With an ionization potential of 35\,eV, double ionized oxygen is typically
seen in only in H{\sc ii} regions of young massive stars (or near AGNs), so the \oiii\,88\,$\mu$m line traces ionized gas (H\,\textsc{ii} regions) and is vastly less dust extinguished than optical
or UV tracers (e.g.,\ \oiii5007\AA). Thus, it  should serve as an excellent tracer of
recent star-formation. The first ionization potential of carbon is only
11.26\,eV, below atomic hydrogen and well below that of double ionized
oxygen, but just above the lowest Lyman-Werner band energy of 11.15\,eV
required to two-step photo-dissociate molecular hydrogen. Furthermore, C\(^+\) is thus
both found in the ionized and neutral gas phases. Recent work using
absorption spectroscopy suggests that \cii can be used to trace the neutral
atomic gas component when metallicity is accounted for \citep{madden20,heintz21}. The lines of \cii and \oiii can thus be used to trace different
parts of the gas associated with the star-forming regions and the diffuse
ISM.  At $z>6,$ a few detections of both lines have been obtained 
\citep[e.g.,][]{inoue16,laporte17,carniani17,hashimoto18,hashimoto19,tamura19,laporte19,harikane20}, revealing a large scatter in the ratios.  While it is not possible with
these data alone to estimate the density and metallicity of the gas, with
sufficient additional data, such estimates may be made from ISM models \citep[e.g.,][]{killi23}.

\begin{figure*}
\setlength{\tabcolsep}{2pt}
\begin{tabular}[t]{@{}lll@{}} 
\includegraphics[trim=0 26 0 0,clip=,width=0.385\textwidth]{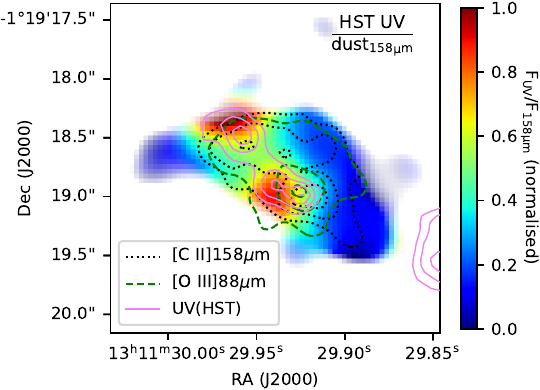}& \includegraphics[trim=52.5 26 0 0,clip=,width=0.307\textwidth]{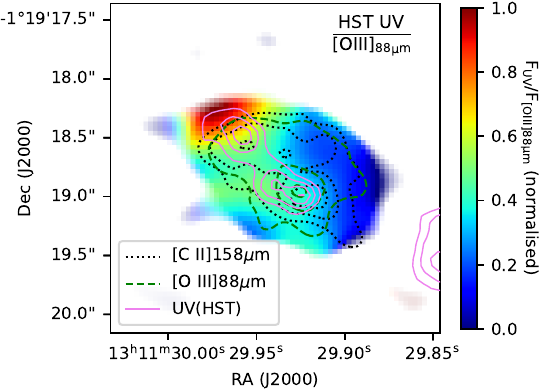}& \includegraphics[trim=52.5 26 0 0,clip=,width=0.307\textwidth]{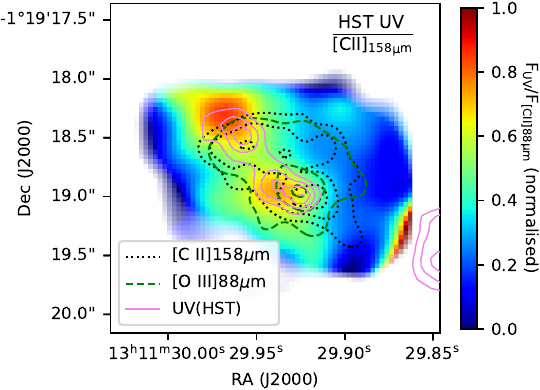}\\[2pt]
\includegraphics[trim=0 26 0 0,clip=,width=0.385\textwidth]{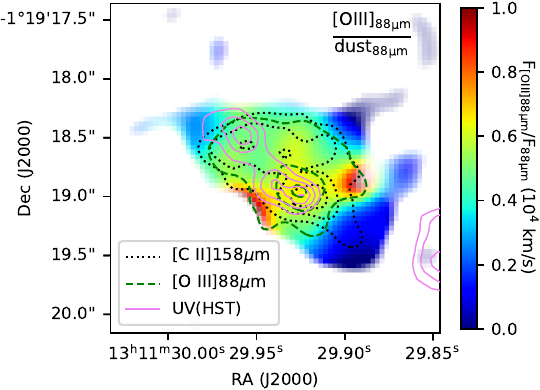}& \includegraphics[trim=52.5 26 0 0,clip=,width=0.307\textwidth]{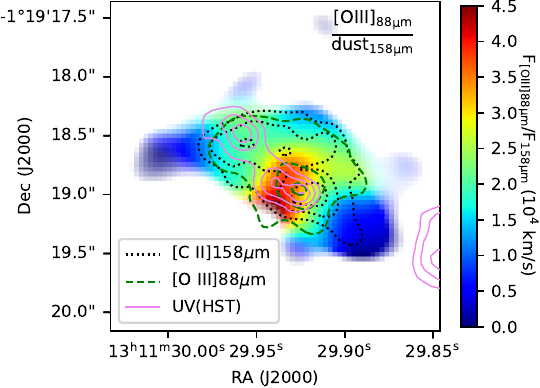}& \includegraphics[trim=52.5 26 0 0,clip=,width=0.307\textwidth]{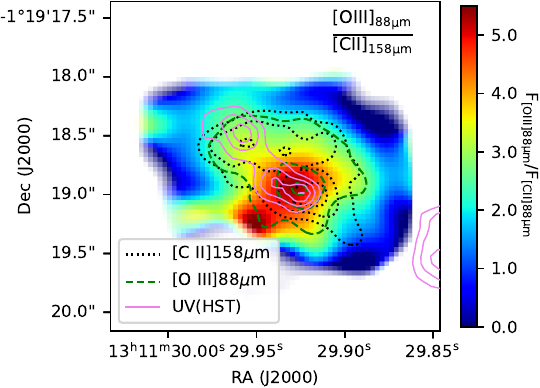}\\[2pt]
\includegraphics[trim=0 0 0 0,clip=,width=0.385\textwidth]{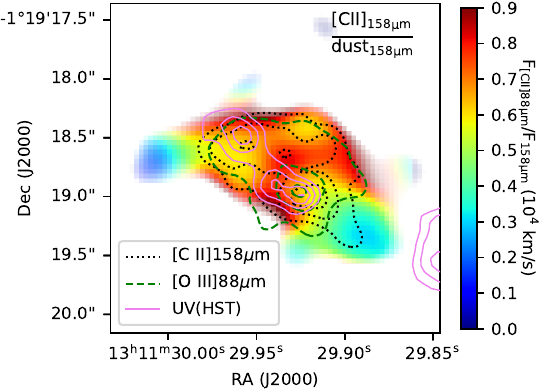}& \includegraphics[trim=52.5 0 0 0,clip=,width=0.307\textwidth]{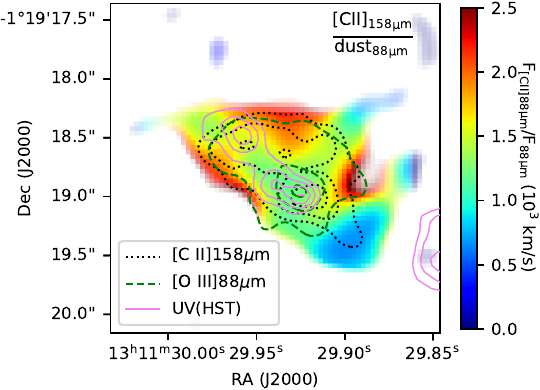}&
\includegraphics[trim=52.5 0 0 0,clip=,width=0.307\textwidth]{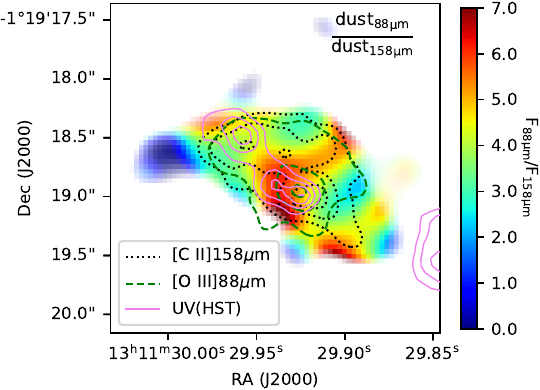}
\end{tabular}
\caption[]{Ratios of the galaxy's UV, \cii158\(\mu\)m, \oiii88\(\mu\)m, and
 dust continuum emission. The images are convolved to the smallest common
 beam:\ the \oiii beam where those data are involved and the \cii beam
 otherwise. Parts of the ratio maps with S/N below 3
 in the denominator are masked white, with increasing transparency from 3 to
 6. The color range is normalized across the S/N \(>6\) part of the map for the panels involving the HST data.
 Contours for the \cii and \oiii line emission and the UV emission are shown
 as dotted black, dashed green, and solid violet  lines.
\label{fig:ratio_plots}}
\end{figure*}

To compare the projected UV emission with the FIR emission in detail, 
we plot in Fig.~\ref{fig:ratio_plots} ratio maps of the \oiii, \cii lines, as
well as the continua near these lines, and the combined UV emission from the
sum of the F125W, F140W, and F160W HST filter images. This gives us
nine ratio maps showing the resolved, normalized line ratios across the galaxy.

Three star-forming tracers are
present: the \oiii line, the UV continuum, and the dust continuum emission.
The UV continuum is susceptible to dust extinction. However, the tracers also
have different overall lifetimes, with the \oiii emission being the most
immediate, requiring 35\,eV photons to ionize the gas, while the UV emission
is longer lived, and the dust heating even longer.
The most notable feature of these ratio maps is that the UV
emission comes only from the eastern side of the galaxy, while the dust and FIR
line emission is much more extended, revealing an entire half of the galaxy
to the west that has no detectable UV emission. Secondly, the \cii emission
also clearly comes from a much more extended area than the other FIR
emission, a point explored in more detail in \citet{akins22}. As noted above, the dusty,
FIR-dominated side of the galaxy seems to correspond to the red and
yellow velocity components in the emission line data.
The yellow component is particularly
unusual in that it is dust-continuum dominated; namely,\ the dust continuum
dominates over all other emission components, including \cii. This component,
however, is mixed in \(88\,\mu\)m and \(158\,\mu\)m continuum emission,
indicating that it is not clearly colder or warmer, for example, than the
dust in the rest of the galaxy. It does, however, also have relatively weak \oiii emission. At first glance, this could suggest that the
oldest star-forming population in the galaxy does in fact reside here. 

While we have argued that the two major components observed in the HST data seem to correspond primarily to the blue and green components identified from the ALMA line emission, it is not a precise match and suggests a greater complexity than what can be resolved with the current data (see Fig.~\ref{fig:cii_HST}).
The central green
component emits strongly in all four tracers, suggesting a young star-forming
knot, while the blue component appears to be more UV dominated, suggesting
that this component is somewhat older and is least dust-rich at its
northern-most extent. 

Finally, the comparison with the high-resolution data of \citet{spilker22} and 
\citet{dessauges19}
show that A1689-zD1 has a very high 
mass surface density. We have argued that the majority of the inferred mass in these surface densities must be stellar mass rather than gas mass, so these results are not quite comparable. However, we expect that a significant fraction of the mass is made up of gas. The high surface density and the range of clump sizes, from unresolved to nearly kiloparsec scale, indicate that A1689-zD1 has some aspects in common with the most massive star-forming galaxies at high-$z$. None of our data so far show any indication of AGN activity in the galaxy \citep[as opposed to the possible interpretation from\footnote{\citet{herreracamus18} found little difference in the \oiii to \cii luminosity ratio between AGNs and starbursts in nearby galaxies. We note our results do not show a central ratio of 7 as otherwise suggested by \citet{Wong2022}.}][]{Wong2022}.

\section{Conclusions}
\label{sect:concl}

We  present GBT band~Q and ALMA band~3, 6, and 8 observations of the gravitationally lensed, epoch of reionization galaxy, A1689-zD1.  The deep ALMA observations combined with the large gravitational magnification, provides some of the deepest high angular resolution data for a $z>7$ galaxy to date, enabling us to resolve the ISM kinematics down to scales of 200pc.  

\begin{itemize}
\item Two emission lines have been detected and identified as \cofe and \oee,
consistent with a redshift of $z=7.1332\pm0.0005$. The spectroscopic redshift is
offset by about 13\,000$\pm7,000$\,km\,s$^{-1}$ from the continuum break identified in previous
X-shooter data.  
\item The line luminosities are $L_{\rm [CII]} = (5.8 \pm 0.1)\times10^{8}$\,L$_\odot$ and $L_{\rm [OIII]} = (1.3 \pm 0.02)\times10^{9}$\,L$_\odot$ after correcting for gravitational magnification.  The line luminosities relative to the SFR are consistent with Local Universe galaxies, suggesting the ISM excitation conditions are normal. 
\item The galaxy system is complex with at least five
individual components.  Based on iterative modeling we estimate the structure and kinematical properties of each component. 
The components appear to be interacting, hinting at a galaxy in its
early assembly. The complex kinematics and morphology do not show signs of a single rotating disk.  The complexity is in agreement with predictions of early galaxy assembly from recent numerical zoom-in simulations.
\item Several components of the galaxy are entirely obscured in the rest-frame UV and might also be similarly obscured in the optical. This could significantly affect the stellar mass estimate, indicating that A1689-zD1 might be significantly more massive than previously believed. Deep rest-frame NIR observations are the most direct route to resolving this question.  
\end{itemize}

Overall, A1689-zD1  still appears to be something of a mystery, with kinematics and gas surface densities that indicate it is in the process of assembly, but the dust and stellar properties  argue for a much older system, with major discrepancies between its dynamical and stellar masses and between its Ly$\alpha$ break and FIR line redshifts, along with a small, compact core, especially in the UV emission; in addition, there is  the vast \cii, and to a lesser extent, dust halo to consider. JWST is certain to resolve at least some of these issues and discrepancies.

The resolved \cii and \oiii kinematics imply a more complex diversity among $z>6$ galaxies.  High-angular-resolution observations of other similar galaxies are needed to improve our understanding of  early assembly  and fully characterize the early evolutionary stages of normal galaxies.

\begin{acknowledgements}
We thank the anonymous referee for their helpful comments and suggestions. 
We acknowledge support from the Nordic ALMA Regional Centre (ARC) node based at Onsala Space Observatory. The Nordic ARC node is funded through Swedish Research Council grant No 2019-00208. 
We thank the Nordic ARC node staff for very helpful discussions. 
KK acknowledges support from the Swedish Research Council (2015-05580) 
and the Knut and Alice Wallenberg Foundation (KAW 2017.0292).  
DW, SF, and NB are supported in part by Independent
Research Fund Denmark grant DFF-7014-00017. The Cosmic Dawn Center is funded by the Danish National Research Foundation under grant number 140.
TB acknowledges support from the Knut and Alice Wallenberg Foundation (KAW 2020.0081).
A.K.I. is supported by NAOJ ALMA Scientific Research grant No.2020-16B.
MJM acknowledges the support of 
the National Science Centre, Poland through the SONATA BIS grant 2018/30/E/ST9/00208.
This research was funded in part by the National Science Centre, Poland, grant number 2023/49/B/ST9/00066. 
This paper makes use of the following ALMA data:
ADS/JAO.ALMA\#2015.1.01406.S and 2017.1.00775.S. ALMA is a partnership of ESO (representing
its member states), NSF (USA) and NINS (Japan), together with NRC
(Canada) and NSC and ASIAA (Taiwan) and KASI (Republic of Korea), in cooperation with the Republic of Chile. The Joint ALMA Observatory is operated by ESO, AUI/NRAO and NAOJ.
\end{acknowledgements}

\bibliographystyle{aa}
\bibliography{aa53229-24}

\end{document}